\documentclass[11pt,a4paper]{article}
\usepackage[utf8]{inputenc}
\usepackage[T1]{fontenc}
\usepackage{amsmath}
\usepackage{amsfonts}
\usepackage{amssymb}
\usepackage{graphicx}
\usepackage{hyperref}
\usepackage{tikz}
\usepackage{algorithm}
\usepackage{algpseudocode}
\usepackage{listings}
\usepackage{xcolor}
\usepackage{booktabs}
\usepackage{multirow}
\usepackage{geometry}
\usepackage{float}

\usetikzlibrary{shapes,arrows,positioning,calc,decorations.pathreplacing}

\definecolor{codegreen}{rgb}{0,0.6,0}
\definecolor{codegray}{rgb}{0.5,0.5,0.5}
\definecolor{codepurple}{rgb}{0.58,0,0.82}
\definecolor{backcolour}{rgb}{0.97,0.97,0.97}

\lstdefinestyle{cppstyle}{
    backgroundcolor=\color{backcolour},
    commentstyle=\color{codegreen},
    keywordstyle=\color{blue},
    numberstyle=\tiny\color{codegray},
    stringstyle=\color{codepurple},
    basicstyle=\small\ttfamily,
    breakatwhitespace=false,
    breaklines=true,
    captionpos=b,
    keepspaces=true,
    numbers=left,
    numbersep=5pt,
    showspaces=false,
    showstringspaces=false,
    showtabs=false,
    tabsize=2,
    frame=single,
    language=C++
}

\hypersetup{
    colorlinks=true,
    linkcolor=blue,
    citecolor=blue,
    urlcolor=blue
}

\title{
    \vspace{-1cm}
    {\Large $\mathcal{L}$}\\[0.5cm]
    \textbf{Lattice: A Post-Quantum Settlement Layer}\\[0.3cm]
    \large A Quantum-Resistant, Peer-to-Peer Electronic Cash System\\
    with Per-Block Difficulty Adjustment and Perpetual Tail Emission
}
\author{
    David Alejandro Trejo Pizzo\\
    \texttt{dtrejopizzo@gmail.com}
}
\date{March 2026 \quad---\quad v0.8.0}

\begin{document}

\maketitle

\begin{abstract}
We present Lattice ($\mathcal{L}$, ticker: LAT), a peer-to-peer electronic cash system designed
as a post-quantum settlement layer for the era of quantum computing. Lattice combines three
independent defense vectors:
\textbf{hardware resilience} through RandomX CPU-only proof-of-work,
\textbf{network resilience} through LWMA-1 per-block difficulty adjustment (mitigating the
Flash Hash Rate vulnerability that affects fixed-interval retarget protocols),
and \textbf{cryptographic resilience} through ML-DSA-44 post-quantum digital signatures
(NIST FIPS 204, lattice-based), enforced exclusively from the genesis block with no
classical signature fallback.
The protocol uses a brief warm-up period of 5,670 fast blocks (53-second target, 25~LAT
reduced reward) for network bootstrap, then transitions permanently to 240-second blocks,
following a 295,000-block halving schedule with a perpetual tail emission floor
of 0.15~LAT per block. Block weight capacity grows in stages (11M$\to$28M$\to$56M) as the
network matures.
The smallest unit of LAT is the \emph{shor}, named after Peter Shor,
where $1~\text{LAT} = 10^8~\text{shors}$.
\end{abstract}

\tableofcontents
\newpage

%% ============================================================================
\section{Introduction}
%% ============================================================================

Bitcoin \cite{nakamoto2008} solved the double-spending problem without a trusted third party.
Its engineering has proven remarkably robust over 15+ years. Lattice builds on Bitcoin's
foundation to explore a different design space---one optimized for three specific concerns:

\begin{enumerate}
    \item \textbf{Hardware accessibility}: SHA-256 mining requires specialized ASICs,
    concentrating participation in industrial operations. Lattice opens mining to any
    computer with a CPU, lowering the barrier to entry.
    \item \textbf{Difficulty responsiveness}: Bitcoin's 2016-block ($\sim$14-day) difficulty
    retarget is stable for large networks but responds slowly to sudden hashrate changes.
    For a new, smaller network where miners may come and go freely, per-block adjustment
    is essential.
    \item \textbf{Quantum readiness}: Shor's algorithm will eventually break ECDSA
    signatures---the timeline is uncertain, but the threat is real. Lattice uses
    post-quantum cryptography from genesis rather than waiting for a migration event.
\end{enumerate}

Lattice addresses all three by building on Bitcoin's proven economic model---the UTXO
structure, the halving schedule, the consensus engine---while making different choices
about mining hardware, difficulty adjustment, and cryptography. Lattice is not an academic
exercise or a research prototype: it is a protocol with a live network, designed
to run for centuries.

\subsection{A Proving Ground for Post-Quantum Transition}

While Lattice is a new network with its own goals, one of its most important contributions
may be as a \textbf{proving ground}. Bitcoin's eventual transition to post-quantum cryptography
is not a question of \emph{if} but \emph{when}---and that transition will be one of the most
complex upgrades in Bitcoin's history: consensus on which algorithm to adopt, a hard or soft
fork, migration of billions of dollars in exposed-key UTXOs, and years of deployment.

Lattice runs this experiment from genesis. By deploying ML-DSA-44 exclusively on a live
proof-of-work network with Bitcoin's UTXO model, halving schedule, and consensus engine,
Lattice produces real-world data on the engineering trade-offs: signature sizes, block
propagation with heavier transactions, storage growth, wallet UX, and cryptographic
performance under production load. Every block mined on Lattice is empirical evidence
that Bitcoin's core developers can use when designing their own post-quantum roadmap.

We consider Bitcoin a marvel of modern engineering. Our goal is not to replace it, but to
contribute---by proving, on a live network, that a path to quantum safety exists and works
at scale.

\subsection{Design Philosophy}

Lattice is a \textbf{settlement layer}---a protocol optimized for secure, final settlement
of value with mathematical guarantees against quantum, hardware, and network attacks.
The design principles are:

\begin{itemize}
    \item \textbf{Sovereignty}: Any computer with a CPU and some RAM is a sovereign node.
    No specialized hardware required. We acknowledge that specialized hardware may eventually
    emerge for any profitable mining algorithm---the goal is not permanent ASIC resistance but
    ensuring that \emph{from day one}, anyone with a commodity CPU can participate on equal
    footing.
    \item \textbf{Adaptive consensus}: The network adjusts difficulty every block, not every two weeks.
    It responds to hashrate changes within hours instead of weeks.
    \item \textbf{Post-quantum only}: ML-DSA-44 signatures are the sole signature scheme. There is
    no ECDSA fallback, no hybrid mode, no transition period. Every transaction from genesis
    block onward is quantum-resistant.
    \item \textbf{Perpetual security}: Tail emission ensures miners are always incentivized to
    secure the network, even centuries from now.
\end{itemize}

\subsection{Complementary, Not Competitive}

Lattice does not compete with Bitcoin. Bitcoin is a marvel of modern engineering---the most
battle-tested distributed system in history. Lattice takes Bitcoin's proven base and extends
it with post-quantum cryptography and CPU-accessible mining, launching a new network from
genesis rather than attempting to modify Bitcoin itself.

The goal is to build a parallel settlement layer that runs alongside Bitcoin.
Bitcoin solved the trust problem. Lattice builds on that solution and adds quantum resistance.
Both networks can coexist---serving different threat models, different time horizons, and
different philosophies about hardware accessibility.

Open protocols exhibit \emph{composability}: each public design decision becomes a
building block that others can adopt, extend, or improve. Ethereum demonstrated this
with DeFi---where every smart contract serves as a public primitive that subsequent
protocols combine into structures no single designer envisioned. The same principle
applies at the protocol layer. Bitcoin's UTXO model, Monero's RandomX, Zawy's LWMA,
and NIST's ML-DSA-44 are all open primitives. Lattice composes them into a configuration
that none of the original authors targeted, yet each component arrives battle-tested.

This is the compound interest of open-source engineering: knowledge accumulates across
projects. Every optimization Monero discovers for RandomX, every cryptanalytic result
NIST publishes on ML-DSA, every difficulty-adjustment refinement the community proposes
for LWMA---all of it flows downstream into Lattice without coordination overhead.

Bitcoin in 2009 designed for a horizon of 2140 (the year block subsidies reach zero).
Lattice designs for a horizon of 2400+. The long-term vision extends beyond centuries---a
store of value secured by mathematics that quantum computers cannot break, running on
the computer you already own. We envision a future where a node runs on your personal
machine: the money travels with you, and the only way to stop it is to confiscate every
computer on Earth. It is a design constraint: every protocol decision made in Lattice is
evaluated against one question, ``will this still work in four centuries?''

\newpage
%% ============================================================================
\section{The Three Pillars}
%% ============================================================================

Lattice combines three independent defense vectors that no existing protocol provides simultaneously:

\begin{figure}[H]
\centering
\begin{tikzpicture}[
    node distance=3.5cm,
    pillar/.style={rectangle, draw, fill=blue!15, text width=7em, text centered,
                   rounded corners, minimum height=4em, font=\small\bfseries},
    desc/.style={text width=10em, text centered, font=\small},
    arrow/.style={thick,->,>=stealth}
]

\node [pillar] (hw) {Hardware\\Resilience};
\node [pillar, right=of hw] (net) {Network\\Resilience};
\node [pillar, right=of net] (crypto) {Cryptographic\\Resilience};

\node [desc, below=0.8cm of hw] (hwd) {RandomX\\CPU + 2GB RAM\\ASIC-resistant};
\node [desc, below=0.8cm of net] (netd) {LWMA-1\\Per-block adjust\\Rapid recovery};
\node [desc, below=0.8cm of crypto] (crd) {ML-DSA-44\\FIPS 204\\PQ-only from genesis};

\draw [thick, decorate, decoration={brace, amplitude=10pt, mirror}]
    ([yshift=-3.5cm]hw.west) -- ([yshift=-3.5cm]crypto.east)
    node[midway, below=12pt, font=\bfseries] {$\mathcal{L}$ --- Lattice};

\end{tikzpicture}
\caption{Three independent defense vectors converge in Lattice.}
\end{figure}
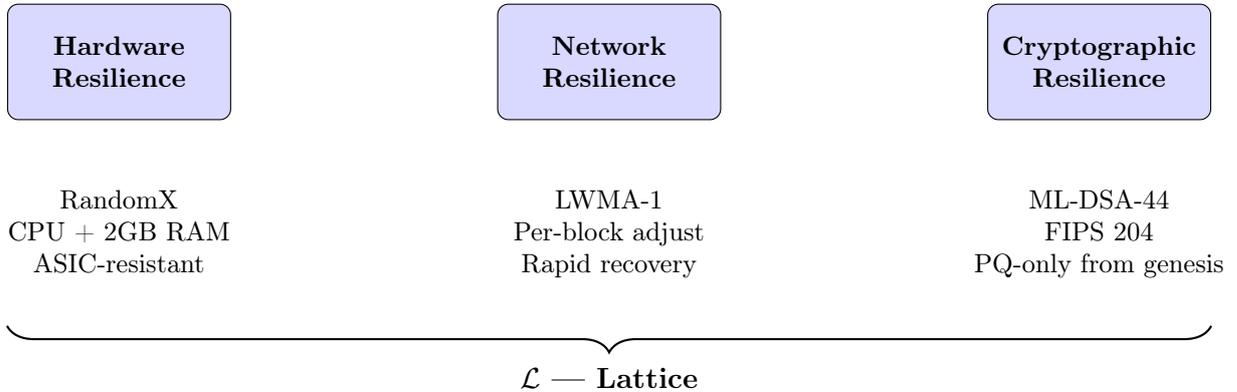

\subsection{Pillar I: Hardware Resilience (RandomX)}
\label{sec:randomx}

RandomX \cite{randomx2019} is a memory-hard proof-of-work algorithm originally developed for
Monero. It delegates security to commodity CPUs and requires 2~GB of RAM per mining thread,
making ASIC fabrication economically unattractive at current network scales. We do not
claim permanent ASIC resistance---history teaches that any sufficiently profitable algorithm
will eventually attract specialized hardware. The goal is to ensure that at network launch
and during early growth, every commodity CPU competes on equal footing.

\textbf{Sovereignty implications}: A nation-state can confiscate a warehouse of ASICs.
It cannot confiscate every laptop on the planet. RandomX makes every general-purpose
computer a potential mining node.

\begin{algorithm}[H]
\caption{RandomX Proof-of-Work}
\begin{algorithmic}
\Require Block header $H$, nonce $n$, seed hash $S$
\Ensure PoW hash $h$

\State $D \leftarrow \texttt{RandomX\_InitDataset}(S)$ \Comment{2~GB dataset from seed}
\State $P \leftarrow \texttt{GenerateProgram}(H \| n)$ \Comment{Random VM program}

\For{$i = 1$ to $8$} \Comment{8 program iterations}
    \State Execute $P$ on scratchpad with random reads from $D$
    \State Update register file and scratchpad state
\EndFor

\State $h \leftarrow \texttt{Blake2b}(\text{register file})$
\Return $h$

\end{algorithmic}
\end{algorithm}

\subsubsection{Memory-Hard Construction}

The 2~GB dataset $D$ is constructed via AES-based expansion:
\begin{equation}
D[i] = \text{AES}_{\text{encrypt}}(D[i-1] \oplus \text{seed}), \quad |D| = 2^{31}~\text{bytes}
\end{equation}

This ensures:
\begin{itemize}
    \item Precomputation of arbitrary dataset portions is infeasible
    \item Memory access patterns are pseudo-random and unpredictable
    \item Time-memory tradeoffs degrade performance superlinearly
    \item Verification requires only a 256~MB cache (``light mode'')
\end{itemize}

\subsubsection{ASIC Resistance Mechanism}

RandomX raises the cost of ASIC development through multiple techniques that force
custom hardware to essentially replicate a general-purpose CPU:
\begin{enumerate}
    \item \textbf{Random code execution}: Each nonce generates a unique program for the RandomX
    virtual machine, preventing fixed-function hardware optimization.
    \item \textbf{Memory latency dependency}: Random reads from the 2~GB dataset create
    memory-latency bottlenecks that cannot be parallelized away.
    \item \textbf{Floating-point arithmetic}: The VM includes IEEE 754 floating-point operations,
    which are native to CPUs but expensive to implement in custom silicon.
    \item \textbf{Branch prediction}: Conditional branches in random programs exploit CPU
    branch prediction units that ASICs would need to replicate.
\end{enumerate}

\subsubsection{CPU Ubiquity: The Decentralization Argument}

There are approximately 4--5 billion CPUs in the world (desktops, laptops, servers).
There are roughly 10--20 million GPUs suitable for mining, and the entire global ASIC
fleet for Bitcoin is concentrated in a few thousand industrial facilities.

\textbf{GPU scarcity in the AI era}: The global supply of high-performance GPUs (NVIDIA
H100, A100, RTX 4090) is now predominantly allocated to artificial intelligence training
and inference workloads. Data centers, cloud providers, and AI companies have absorbed
the majority of production capacity, creating sustained shortages and multi-month wait
times for GPU procurement. This structural demand will persist for at least the next
decade as AI model sizes and training compute continue to scale. For a new network
launching in 2026, designing around GPU mining would mean competing for hardware against
the most capital-intensive industry in history. CPU mining sidesteps this entirely---CPUs
are abundant, general-purpose, and not subject to AI-driven scarcity.

\begin{center}
\begin{tabular}{lrl}
\toprule
\textbf{Hardware} & \textbf{Est.\ Global Units} & \textbf{Mining Access} \\
\midrule
CPUs (desktop/laptop/server) & $\sim$4--5 billion & Anyone \\
GPUs (mining-capable) & $\sim$10--20 million & Enthusiasts, farms \\
Bitcoin ASICs & $\sim$3--5 million & Industrial operators \\
\bottomrule
\end{tabular}
\end{center}

This distribution matters---not because billions of CPUs \emph{will} mine, but because
they \emph{can}. In practice, most CPU owners will not mine, just as most Bitcoin node
operators do not run mining ASICs. The realistic miner population will consist of
enthusiasts, small server operators, cloud instances, and some botnets---similar to
Monero's ecosystem. The difference from ASIC mining is not that ``everyone mines'' but
that the \emph{barrier to entry is zero}: anyone who wants to participate can, without
purchasing specialized hardware or competing against industrial operations.

\textbf{Cloud mining in early stages}: Critics note that RandomX allows mining on rented
cloud infrastructure (AWS, GCP, Azure). In the early network, this is beneficial: every
cloud instance mining Lattice is also a full node validating the chain, increasing both
hashrate \emph{and} node count simultaneously. However, we recognize that cloud mining
concentration is not desirable long-term---a network dependent on a few cloud providers
inherits their policy risks (see Scenario 7 in Section~\ref{sec:red-team}).

In practice, the barrier to cloud mining is higher than it appears: a user must provision
a VPS, install Docker, clone a repository, configure networking, and manage the instance---a
process that requires meaningful technical skill. For someone with that skill level, running
a node on their own hardware (a desktop, a used server, a Mac Mini) is simpler and cheaper.
The protocol is designed so that mining on hardware you already own is always the
path of least resistance.

\subsubsection{Known Limitation: Botnets}
\label{sec:botnets}

RandomX's low hardware barrier means that any CPU can mine, including compromised machines in
botnets. This is an inherent trade-off of CPU-friendly mining: the same property that enables
sovereignty (any computer can mine) also means that illicitly controlled computers can mine.
This limitation applies equally to all CPU-mineable cryptocurrencies (notably Monero) and is
not unique to Lattice.

\textbf{Quantitative analysis}: We model the threat concretely.

A typical consumer CPU (Intel i5-12400, AMD Ryzen 5 5600) achieves $\sim$5,000--8,000 H/s
on RandomX. A compromised machine in a botnet operates at reduced efficiency due to thermal
throttling, shared CPU time, and the requirement to remain undetected:

\begin{center}
\begin{tabular}{lrrl}
\toprule
\textbf{Botnet Size} & \textbf{Effective H/s} & \textbf{Equiv.\ Dedicated Nodes} & \textbf{Notes} \\
\midrule
10,000 PCs & $\sim$20~MH/s & $\sim$3,000 & Small botnet \\
100,000 PCs & $\sim$200~MH/s & $\sim$30,000 & Moderate (Zeus-class) \\
1,000,000 PCs & $\sim$2~GH/s & $\sim$300,000 & Large (Mirai-class) \\
5,000,000 PCs & $\sim$10~GH/s & $\sim$1,500,000 & Nation-state scale \\
\bottomrule
\end{tabular}
\end{center}

The effective hashrate per bot assumes $\sim$2,000~H/s average (40\% of a dedicated node)
due to: (a)~the 2~GB RAM requirement excludes most IoT devices and phones, (b)~shared CPU
with legitimate user workloads, (c)~thermal throttling on laptops, (d)~intermittent
availability (machines sleep, reboot, get cleaned).

\textbf{Detection economics}: Running RandomX at meaningful rates consumes 65--125W per core
and saturates at least one thread continuously. Users notice degraded performance; enterprise
antivirus flags sustained 100\% CPU; electricity costs increase on the victim's bill. The
cost of maintaining a 1M-machine botnet for mining is estimated at \$50K--\$200K/month in
C2 infrastructure, exploit maintenance, and attrition (machines get cleaned at 5--15\%/month).

\textbf{Economic asymmetry}: The most important property of botnet mining is not hashrate
efficiency but \emph{cost structure}. A botnet operator's marginal electricity cost is zero---the
victim pays the power bill. An honest miner pays \$0.10--\$0.15/kWh. This creates an asymmetry:
botnet hashrate is profitable at \emph{any} token price above zero, while honest miners require
a minimum price to cover electricity. If botnet hashrate grows, it pushes difficulty upward,
squeezing honest miners whose costs are real. In the worst case, this dynamic displaces
legitimate miners at the margin.

\textbf{Monero precedent}: Monero has used RandomX since November 2019. Botnet mining
has been documented \cite{moser2018} (e.g., Smominru, estimated 500K+ machines) but has never exceeded
$\sim$2--5\% of Monero's total network hashrate. The network has remained stable and secure
despite botnet presence for 6+ years. The empirical evidence suggests that the economic
asymmetry, while real, has not been sufficient to displace honest miners in practice---likely
because botnet attrition rates (5--15\%/month), the 2~GB RAM filter, and detection risk impose
hidden costs that narrow the gap. Lattice inherits this empirical resilience, but acknowledges
that botnet economics remain an inherent and unresolved trade-off of CPU-friendly mining.

\subsubsection{Hardware Requirements}

\begin{center}
\begin{tabular}{lcc}
\toprule
\textbf{Component} & \textbf{Minimum} & \textbf{Recommended} \\
\midrule
CPU & x86-64 / ARM64 (4+ cores) & Modern 8+ cores \\
RAM & 16~GB & 32~GB+ \\
Storage & 256~GB SSD & 500~GB+ SSD \\
Network & 1~Mbps & 10~Mbps+ \\
\bottomrule
\end{tabular}
\end{center}

\textbf{Why 16~GB RAM minimum}: We created an install script for Lattice that runs two
Docker containers simultaneously (mining worker + RPC node). This was done to make it
easier for anyone with basic coding knowledge to start a node. Each container requires
the 2~GB RandomX dataset in memory, plus overhead for the Bitcoin Core process, UTXO
cache, and the operating system. Systems with less than 16~GB of total RAM may
experience out-of-memory crashes due to Docker's default memory allocation limits.

\textbf{Reference deployment}: A Lattice node can run on an Apple Mac Mini (M4, 16~GB
RAM, 256~GB SSD)---a consumer desktop machine. This configuration runs a full mining node
and RPC node simultaneously via Docker without performance issues. No server-grade hardware,
GPU, or data center infrastructure is required. See Section~\ref{sec:blockweight} for
storage growth projections at different adoption levels.

\subsection{Pillar II: Network Resilience (LWMA-1)}
\label{sec:lwma}

Bitcoin retargets difficulty every 2016 blocks ($\sim$14 days). This creates a
vulnerability: the \textbf{Flash Hash Rate} problem:

\begin{enumerate}
    \item A large mining pool (or state actor) joins the network with significant hashrate
    \item Difficulty adjusts upward to match the new total hashrate
    \item The attacker suddenly withdraws all hashrate
    \item The remaining miners face elevated difficulty
    \item Blocks take hours or days instead of minutes
    \item Under Bitcoin's retarget algorithm, the network operates at degraded performance
    until the next 2016-block retarget---potentially up to 14 days per cycle
\end{enumerate}

With Bitcoin's 4$\times$ cap on difficulty adjustment per retarget period, recovery from
a large hashrate withdrawal is slow but eventual. A 10$\times$ hashrate drop would require
multiple retarget periods to fully recover, resulting in weeks to months of degraded block times.

\textbf{Note}: Bitcoin has never experienced this attack in practice, partly because its
hashrate is dominated by specialized ASICs with high sunk costs that disincentivize
rapid withdrawal. For CPU-mineable chains like Lattice, where miners can freely redirect
hashrate to other uses, per-block adjustment is essential.

\subsubsection{LWMA-1: Per-Block Difficulty Adjustment}

Lattice implements zawy12's Linear Weighted Moving Average (LWMA-1) algorithm
\cite{zawy2017lwma}, which adjusts difficulty \textbf{every single block} based on a
weighted average of the most recent 120 solve times:

\begin{equation}
\text{nextTarget} = \frac{\text{sumTarget}}{N} \cdot \frac{\sum_{i=1}^{N} i \cdot t_i}{k}
\end{equation}

where:
\begin{itemize}
    \item $N = 120$ (window size, covering $\sim$8 hours at 240s target spacing)
    \item $t_i = \text{timestamp}[i] - \text{timestamp}[i-1]$ is the solve time for block $i$
    \item $k = \frac{N(N+1)}{2} \cdot T$ is the expected weighted sum at target spacing $T$
    \item Recent blocks have higher weight (block $i$ is weighted by $i$)
\end{itemize}

\textbf{Safety bounds}:
\begin{itemize}
    \item \textbf{Solve time clamp}: $|t_i| \leq 6T$ (prevents timestamp manipulation)
    \item \textbf{Floor}: $\sum i \cdot t_i \geq k/10$ (prevents extreme difficulty spikes---
    difficulty cannot increase more than 10$\times$ in a single window)
    \item \textbf{Ceiling}: Target never exceeds \texttt{powLimit}
\end{itemize}

\begin{lstlisting}[caption={LWMA-1 Core Implementation (pow.cpp)}, label=lst:lwma]
const int N = 120; // LWMA window
const int64_t k = N * (N + 1) * T / 2;

arith_uint256 sumTarget;
int64_t weightedSolveTimeSum = 0;

for (int i = N; i > 0; i--) {
    int64_t solvetime = pindex->GetBlockTime()
                      - pindexPrev->GetBlockTime();

    // Clamp solve time to prevent timestamp manipulation
    if (solvetime > 6 * T)  solvetime = 6 * T;
    if (solvetime < -(6*T)) solvetime = -(6 * T);

    weightedSolveTimeSum += solvetime * i;
    sumTarget += target / (uint32_t)N;
    pindex = pindexPrev;
}

// Floor: prevent extreme difficulty spikes
if (weightedSolveTimeSum < k / 10)
    weightedSolveTimeSum = k / 10;

// Safe overflow handling for uint256 arithmetic
if ((sumTarget >> 192) > arith_uint256(0)) {
    nextTarget = sumTarget / (uint32_t)k;
    nextTarget *= (uint32_t)weightedSolveTimeSum;
} else {
    nextTarget = sumTarget * (uint32_t)weightedSolveTimeSum;
    nextTarget /= (uint32_t)k;
}
\end{lstlisting}

\subsubsection{LWMA vs. Bitcoin's Retarget: Comparison}

\begin{center}
\begin{tabular}{lcc}
\toprule
\textbf{Property} & \textbf{Bitcoin (2016-block)} & \textbf{Lattice (LWMA-1)} \\
\midrule
Adjustment frequency & Every $\sim$14 days & Every block \\
Response to 10$\times$ hash drop & Multiple retarget periods & $\sim$8 hours recovery \\
Timestamp manipulation & Vulnerable & Clamped to $\pm 6T$ \\
Difficulty oscillation & Moderate & Smoothed by linear weights \\
\bottomrule
\end{tabular}
\end{center}

\subsection{Pillar III: Cryptographic Resilience (ML-DSA-44)}
\label{sec:mldsa}

Quantum computing poses two distinct threats to cryptocurrency:

\begin{enumerate}
    \item \textbf{Shor's Algorithm} \cite{shor1997}: Breaks ECDSA/EdDSA signatures
    in polynomial time $O(n^3)$, enabling private key extraction from public keys.
    Requires $\sim$4,000 stable logical qubits. This is the \textbf{primary threat}.

    \item \textbf{Grover's Algorithm} \cite{grover1996}: Provides quadratic speedup
    $O(\sqrt{N})$ for hash inversion, reducing effective security from 256 to 128 bits.
    Affects \textbf{all} PoW algorithms equally (SHA-256, RandomX, etc.).
    Addressed by difficulty adjustment when it becomes practical.
\end{enumerate}

\textbf{Industry timelines} (as of 2025):

\begin{center}
\small
\begin{tabular}{llrl}
\toprule
\textbf{Organization} & \textbf{Target} & \textbf{Year} & \textbf{Source} \\
\midrule
IBM & 100,000 qubits & 2033 & \cite{ibmroadmap2023} \\
Google & Error-corrected logical qubits & 2029 & \cite{google2024willow} \\
NIST & ``Migrate away from ECDSA now'' & 2035 & \cite{nistir8547} \\
NSA/CNSA & Quantum-safe required for classified & 2030 & \cite{cnsa2022} \\
BSI (Germany) & PQ transition deadline & 2030 & \cite{bsi2024} \\
\bottomrule
\end{tabular}
\end{center}

\textbf{Critical point}: The quantum timeline is uncertain. Cryptographically relevant
quantum computers may arrive in 2035, 2050, or 2080---or later. Lattice does not bet on
any specific date. The protocol is designed so that the quantum threat timeline is
irrelevant: whether it is 10 years away or 30, every transaction is already protected.

\textbf{Quantum computers vs.\ quantum annealers}: Public discussion often conflates
\emph{gate-based quantum computers} (IBM, Google, PsiQuantum) with \emph{quantum annealers}
(D-Wave). This distinction matters: Shor's algorithm requires a universal gate-based quantum
computer with error correction. Quantum annealers solve optimization problems and
\textbf{cannot run Shor's algorithm}. Current claims of ``quantum supremacy'' in 2026 from annealer
vendors are irrelevant to cryptographic security. The threat to ECDSA comes exclusively from
gate-based machines, which are further from practical deployment but advancing rapidly.

\textbf{Timeline uncertainty}: It is possible that practical gate-based quantum computers
capable of breaking ECDSA will take longer than industry roadmaps suggest. Error correction
remains the bottleneck, and extrapolating from current qubit counts to millions of logical
qubits involves significant engineering unknowns. However, the pace of progress is
accelerating: advances in artificial intelligence are increasingly applied to quantum error
correction, qubit design, and materials science, potentially compressing timelines that
were previously measured in decades. Lattice acknowledges the uncertainty in both
directions---but argues that the cost of being prepared (larger signatures) is far lower
than the cost of being unprepared (total loss of funds).

Even if quantum computers never materialize at the scale needed to break ECDSA, Lattice
loses nothing: ML-DSA-44 provides 128-bit security against classical adversaries as well.

\textbf{The ``harvest now, decrypt later'' threat} \cite{mosca2018}: Intelligence agencies
are known to store encrypted communications today for future quantum decryption. In
cryptocurrency, every transaction that reveals a public key (Bitcoin's default behavior)
creates a ``harvest now, spend later'' vulnerability: an adversary records exposed public
keys and waits for a quantum computer to extract private keys. Bitcoin has an estimated
5--10 million BTC in addresses with exposed public keys \cite{stewart2018}. Lattice will have zero by design.

\subsubsection{ML-DSA-44: NIST FIPS 204}

Lattice implements ML-DSA-44 (Module-Lattice Digital Signature Algorithm) \cite{nist2024pqc},
the NIST standard for post-quantum digital signatures. Unlike hybrid approaches that combine
classical and post-quantum schemes \cite{kiktenko2018}, Lattice enforces ML-DSA-44 exclusively---there
is no classical fallback to compromise. The algorithm's security is based
on the hardness of the \textbf{Module Learning With Errors (MLWE)} problem---a lattice-based
mathematical problem for which no efficient quantum algorithm exists \cite{regev2009}.

\begin{center}
\begin{tabular}{lcc}
\toprule
\textbf{Property} & \textbf{ECDSA (Bitcoin)} & \textbf{ML-DSA-44 (Lattice)} \\
\midrule
Security basis & Elliptic curve DLP & Module-LWE (lattice) \\
Quantum resistant & No (Shor's) & Yes (no known quantum attack) \\
Public key size & 33 bytes & 1,312 bytes \\
Signature size & 72 bytes & 2,420 bytes \\
NIST standard & --- & FIPS 204 \\
Security level & 128-bit classical & 128-bit post-quantum (Category 2) \\
\bottomrule
\end{tabular}
\end{center}

\subsubsection{PQ-Only Architecture: No ECDSA Fallback}

Lattice enforces ML-DSA-44 as the \textbf{sole} signature algorithm from the genesis block.
ECDSA is disabled at the consensus level:

\begin{lstlisting}[caption={PQ-Only Signature Verification (interpreter.cpp, simplified)}]
case OP_CHECKSIG:
case OP_CHECKSIGVERIFY: {
    valtype& vchPubKey = stacktop(-1);
    valtype& vchSig    = stacktop(-2);

    if (IsPQCPubKey(vchPubKey)) {
        // ML-DSA-44 post-quantum verification
        fSuccess = VerifyMLDSA44Signature(
            vchSig, vchPubKey, sighash);
    } else {
        // ECDSA is disabled at consensus level.
        // Non-PQC public keys are rejected.
        fSuccess = false;
    }
    break;
}
\end{lstlisting}

\textbf{Rationale}: A hybrid scheme (ECDSA + ML-DSA-44) would contradict the protocol's
security model. If ECDSA is accepted, an attacker with a quantum computer can forge ECDSA
signatures, making the ML-DSA-44 layer irrelevant. A post-quantum chain must be
post-quantum \emph{exclusively}, or it is not post-quantum at all. This means Lattice cannot interoperate with Bitcoin wallets or reuse Bitcoin private keys.

\subsubsection{Lattice}

The project name is not accidental. ML-DSA-44's security derives from lattice-based
cryptography---the mathematical structure of high-dimensional lattices in which the
Shortest Vector Problem (SVP) and Learning With Errors (LWE) are believed to be hard
even for quantum computers. Lattice the protocol is protected by lattices the mathematics.

\newpage
%% ============================================================================
\section{Post-Quantum Cryptography: Full PQC from Genesis}
\label{sec:pqc}
%% ============================================================================

Lattice is a \textbf{Full PQC} protocol: every cryptographic operation---from key generation
through transaction signing and verification---uses exclusively post-quantum algorithms.
There is no classical signature fallback, no hybrid mode, and no transition period. This
section explains the landscape of post-quantum signature schemes, why ML-DSA-44 was selected,
and how its security compares to Bitcoin's ECDSA.

\subsection{The Post-Quantum Signature Landscape}

NIST's Post-Quantum Cryptography Standardization Process (2016--2024) evaluated dozens of
candidate signature schemes over 8 years of public cryptanalysis. The finalists and their
trade-offs are:

\begin{center}
\small
\begin{tabular}{lrrrcl}
\toprule
\textbf{Scheme} & \textbf{Pubkey} & \textbf{Signature} & \textbf{Security} & \textbf{NIST} & \textbf{Basis} \\
& \textbf{(bytes)} & \textbf{(bytes)} & \textbf{(bits)} & \textbf{Status} & \\
\midrule
\textbf{ML-DSA-44} & 1,312 & 2,420 & 128 PQ & FIPS 204 & Module-LWE lattice \\
ML-DSA-65 & 1,952 & 3,309 & 192 PQ & FIPS 204 & Module-LWE lattice \\
ML-DSA-87 & 2,592 & 4,627 & 256 PQ & FIPS 204 & Module-LWE lattice \\
FALCON-512 & 897 & 666 & 128 PQ & FIPS 206 & NTRU lattice \\
FALCON-1024 & 1,793 & 1,280 & 256 PQ & FIPS 206 & NTRU lattice \\
SLH-DSA-128s & 32 & 7,856 & 128 PQ & FIPS 205 & Hash-based \\
SLH-DSA-128f & 32 & 17,088 & 128 PQ & FIPS 205 & Hash-based \\
XMSS & 64 & 2,500 & 128 PQ & RFC 8391 & Hash-based (stateful) \\
\midrule
ECDSA (secp256k1) & 33 & 72 & 128 classical & --- & Elliptic curve DLP \\
\bottomrule
\end{tabular}
\end{center}

\subsection{Why ML-DSA-44 (Dilithium2)}

The selection of ML-DSA-44 over other candidates was driven by five criteria:

\begin{enumerate}
    \item \textbf{NIST standardization}: ML-DSA-44 is the primary NIST standard for
    post-quantum signatures (FIPS 204, finalized August 2024). It received the most
    extensive public cryptanalysis of any PQ signature scheme---8 years of review by
    the global cryptographic community. FALCON (FIPS 206) is also standardized but
    arrived later and with more implementation complexity.

    \item \textbf{Stateless design}: ML-DSA-44 is \textbf{stateless}---signing does not
    require tracking which keys have been used. This is critical for cryptocurrency
    wallets where address reuse, backup restoration, and multi-device access are common.
    XMSS is \emph{stateful}: reusing a one-time key reveals the private key, making it
    fundamentally unsuitable for a UTXO model where users may sign multiple transactions
    from the same address. SLH-DSA (SPHINCS+) is stateless but produces enormous
    signatures (7--17~KB), making it impractical for on-chain use.

    \item \textbf{Balanced size profile}: ML-DSA-44 offers the best trade-off between
    public key size (1,312 bytes), signature size (2,420 bytes), and verification speed.
    FALCON-512 has smaller signatures (666 bytes) and smaller keys (897 bytes) but requires
    constant-time floating-point arithmetic during signing---a notoriously difficult
    implementation challenge that has produced side-channel vulnerabilities in multiple
    libraries. FALCON's signing is also 3--5$\times$ slower than ML-DSA-44's.

    \item \textbf{Implementation simplicity}: ML-DSA-44 uses only integer arithmetic
    (no floating-point), making constant-time implementation straightforward. The
    reference implementation is approximately 2,000 lines of C. FALCON requires
    careful floating-point sampling from a discrete Gaussian distribution---an operation
    that is easy to get wrong and hard to audit. For a security-critical protocol that
    must run correctly on every platform (x86, ARM, RISC-V), implementation simplicity
    is a first-order concern.

    \item \textbf{Performance}: ML-DSA-44 key generation, signing, and verification are
    all fast on commodity CPUs:

    \begin{center}
    \begin{tabular}{lrrr}
    \toprule
    \textbf{Operation} & \textbf{ML-DSA-44} & \textbf{FALCON-512} & \textbf{ECDSA} \\
    \midrule
    Key generation & $\sim$0.05 ms & $\sim$8.0 ms & $\sim$0.04 ms \\
    Signing & $\sim$0.15 ms & $\sim$0.40 ms & $\sim$0.05 ms \\
    Verification & $\sim$0.05 ms & $\sim$0.05 ms & $\sim$0.15 ms \\
    \bottomrule
    \end{tabular}
    \end{center}

    Benchmarks on Intel i7-12700 (single core). ML-DSA-44 verification is actually
    \emph{faster} than ECDSA verification---a counter-intuitive result that arises
    because ML-DSA-44 verification involves only matrix-vector multiplication over
    small integers, while ECDSA requires expensive elliptic curve point multiplication.
\end{enumerate}

\textbf{Why not FALCON?} Despite FALCON's size advantage (666~B vs.\ 2,420~B signatures),
the implementation risk is unacceptable for a genesis-launch protocol. FALCON signing
requires sampling from a discrete Gaussian distribution using floating-point arithmetic
that must be constant-time to prevent side-channel attacks. Multiple implementations
have been found vulnerable \cite{nist2024pqc}. Lattice prioritizes correctness and
auditability over byte efficiency. If FALCON matures and proves itself in production
deployments, a future soft fork could add it as an additional signature type.

\textbf{Why not SLH-DSA (SPHINCS+)?} SLH-DSA's security is based on hash functions
alone---arguably the most conservative security assumption possible. However, its
signatures are 7,856--17,088 bytes, making on-chain transactions 3--7$\times$ larger than
ML-DSA-44. At 56M block weight, this would reduce throughput to 2--5~tx/s, making the
network impractical for settlement use.

\subsection{Full PQC Architecture: Key Generation to Verification}

Lattice's PQ-only architecture means every layer of the cryptographic stack is
post-quantum:

\begin{enumerate}
    \item \textbf{Key generation}: The \texttt{pq\_getnewaddress} RPC generates an
    ML-DSA-44 keypair via liboqs (Open Quantum Safe). The private key (2,560 bytes)
    and public key (1,312 bytes) are combined into a 3,872-byte secret key (7,744 hex
    characters) stored in a JSON wallet file. No BIP39 seed phrases, no ECDSA derivation
    paths, no HD wallet hierarchy---the keypair is the identity.

    \item \textbf{Address derivation}: The public key is hashed with Hash160
    (RIPEMD160(SHA256)) and encoded as a Base58Check address with Lattice's version
    byte 48, producing addresses that start with \texttt{L} (e.g., \texttt{LKHDz\ldots}).
    The \texttt{pq\_importkey} RPC imports the public key into the node's wallet as a
    \texttt{raw()} descriptor, enabling UTXO tracking for that address.

    \item \textbf{Transaction signing}: When spending UTXOs, the wallet loads the private
    key from the JSON wallet file and signs the transaction hash using ML-DSA-44's
    deterministic signing algorithm. The 2,420-byte signature and 1,312-byte public key
    are placed in the P2PKH scriptSig ($\sim$3,740~bytes total). Future versions may
    migrate to SegWit witness structures to benefit from the 4$\times$ weight discount.

    \item \textbf{Consensus verification}: Every node verifies every signature using
    ML-DSA-44. ECDSA signatures are \textbf{rejected at the consensus level}---the
    \texttt{OP\_CHECKSIG} opcode only accepts PQ public keys. This is enforced from
    the genesis block with no transition period.
\end{enumerate}

\subsection{Security Comparison: ML-DSA-44 vs.\ Bitcoin's ECDSA}

The fundamental question: does Lattice's post-quantum cryptography make keys
\emph{harder} or \emph{easier} to break compared to Bitcoin?

\subsubsection{Classical Security (No Quantum Computer)}

Against a classical (non-quantum) adversary, both schemes provide 128-bit security:

\begin{center}
\begin{tabular}{lcc}
\toprule
\textbf{Property} & \textbf{ECDSA (secp256k1)} & \textbf{ML-DSA-44} \\
\midrule
Security level & 128-bit & 128-bit \\
Best classical attack & $2^{128}$ (Pollard's rho) & $2^{299}$ (BKZ sieving) \\
Security margin & $2^{128}/2^{128} = 1$ & $2^{299}/2^{128} = 2^{171}$ \\
Attack cost (\$ estimate) & $> 10^{25}$ USD & $> 10^{76}$ USD \\
\bottomrule
\end{tabular}
\end{center}

\textbf{Key insight}: Against classical attacks, ML-DSA-44 is \textbf{astronomically
more secure} than ECDSA. The best known classical attack against ML-DSA-44 requires
$2^{299}$ operations---a margin of $2^{171}$ above the 128-bit security target. ECDSA's
128-bit security is tight: the best attack (Pollard's rho on elliptic curves) achieves
exactly $2^{128}$ operations. If a breakthrough reduces the security of lattice problems
by even 100 bits, ML-DSA-44 would still provide 199-bit security---far above the
cryptographic safety threshold.

\subsubsection{Quantum Security (With Quantum Computer)}

This is where the comparison becomes stark:

\begin{center}
\begin{tabular}{lcc}
\toprule
\textbf{Property} & \textbf{ECDSA (secp256k1)} & \textbf{ML-DSA-44} \\
\midrule
Quantum security & \textbf{0 bits} & 128 bits \\
Best quantum attack & Shor's: $O(n^3)$ poly-time & $2^{271}$ (quantum sieving) \\
Qubits required to break & $\sim$4,000 logical & $> 10^{40}$ (infeasible) \\
Key extraction from pubkey & Yes (polynomial time) & No known method \\
``Harvest now, break later'' & Vulnerable & Not applicable \\
\bottomrule
\end{tabular}
\end{center}

\textbf{ECDSA under quantum attack}: Shor's algorithm factors the discrete logarithm
problem in polynomial time $O(n^3)$, where $n$ is the key size in bits. For secp256k1
(256-bit keys), a quantum computer with $\sim$4,000 stable logical qubits can extract
the private key from any exposed public key in hours. The security drops from 128 bits
to \textbf{effectively zero}. Every Bitcoin address that has ever sent a transaction
(exposing its public key) becomes vulnerable. An estimated 5--10 million BTC ($\sim$25\%
of supply) sits in addresses with exposed public keys \cite{stewart2018}.

\textbf{ML-DSA-44 under quantum attack}: The best known quantum attack against
Module-LWE uses quantum sieving algorithms with cost $2^{271}$---still astronomically
beyond any feasible computation. No polynomial-time quantum algorithm is known for
lattice problems. The mathematical structure of lattice problems (finding short vectors
in high-dimensional spaces) does not admit the same algebraic shortcuts that Shor's
algorithm exploits in number theory. NIST's 8-year evaluation process specifically
assessed quantum resilience and concluded that ML-DSA-44 provides 128-bit
post-quantum security (Category 2).

\subsubsection{Summary: Security Trade-Off}

\begin{center}
\begin{tabular}{lcc}
\toprule
\textbf{Scenario} & \textbf{Bitcoin (ECDSA)} & \textbf{Lattice (ML-DSA-44)} \\
\midrule
Classical adversary (today) & 128-bit (adequate) & 128-bit (massive margin) \\
Quantum adversary (future) & \textbf{Broken} & 128-bit (secure) \\
Key size penalty & 33 bytes & 1,312 bytes ($40\times$) \\
Signature size penalty & 72 bytes & 2,420 bytes ($34\times$) \\
Verification speed & $\sim$40K/s & $\sim$20K/s ($2\times$ slower) \\
\bottomrule
\end{tabular}
\end{center}

The trade-off is explicit: Lattice pays a $\sim$34$\times$ signature size penalty and
$\sim$2$\times$ verification cost in exchange for immunity to the most powerful known
computational threat. Storage is cheap (\$0.05/GB) and getting cheaper. Broken
cryptography is not cheap---it is catastrophic and irreversible. The entire history of
a chain secured by broken cryptography becomes retroactively vulnerable: funds can be
stolen, transactions can be forged, and the chain's integrity collapses.

Lattice chooses to pay the storage cost today rather than risk cryptographic collapse
tomorrow. This is not a hedge---it is an engineering decision. A protocol designed for
centuries cannot be built on mathematics that has a known expiration date.

\newpage
%% ============================================================================
\section{Warm-Up Period and Block Timing}
\label{sec:warmup}
%% ============================================================================

Lattice uses a brief warm-up period at network launch, followed by a permanent block time
for the rest of the chain's lifetime.

\subsection{Warm-Up: Network Bootstrap (Blocks 0--5,669)}

\begin{itemize}
    \item \textbf{Block time}: 53 seconds (target)
    \item \textbf{Duration}: $\sim$83.5 hours (3.5 days)
    \item \textbf{Blocks}: 5,670
    \item \textbf{Block reward}: 25~LAT (reduced from 50~LAT to limit early mining advantage)
    \item \textbf{LAT mined}: 141,750 (at 25~LAT/block)
    \item \textbf{Purpose}: Rapid network bootstrap with fair distribution
\end{itemize}

The warm-up serves a practical purpose: when a new proof-of-work chain launches with zero
hashrate, the first blocks must come quickly enough to establish connectivity, distribute
initial coins, and allow LWMA-1 to accumulate enough history (120 blocks) to function properly.
Without a warm-up, the initial 240-second target combined with near-zero difficulty would
still produce fast blocks, but the explicit warm-up ensures predictable bootstrap behavior
regardless of initial hashrate.

\textbf{Reduced warm-up reward}: During the warm-up, block rewards are capped at 25~LAT
(half the normal 50~LAT subsidy). This limits the advantage of early miners who participate
when difficulty is lowest. The warm-up produces 141,750~LAT total---less than 0.5\% of the
eventual supply---ensuring fair distribution without creating a significant early-miner premium.

At block 5,670, difficulty resets to \texttt{powLimit} to allow LWMA to recalibrate for
the new 240-second target spacing. This transition takes approximately 120 blocks
($\sim$8 hours) to stabilize.

\textbf{Instamine resistance}: LWMA-1 is active from block 1, not just after the warm-up.
If a cloud burst attempts to accelerate the warm-up (e.g., renting massive CPU capacity
to mine all 5,670 blocks in hours instead of 83.5 hours), LWMA-1 responds by increasing
difficulty within the first 120 blocks, throttling the acceleration. The attacker would
need to sustain growing hashrate against rising difficulty---an expensive proposition
for 5,670 blocks of a network with zero economic value at launch. The 25~LAT reward cap
further reduces the incentive for such attacks by halving the potential reward.

\textbf{Orphan risk during warm-up}: The 53-second block time combined with ML-DSA-44's
larger block sizes increases orphan risk during the warm-up. This is acceptable because
(a) the warm-up lasts only $\sim$83.5 hours, (b) transaction volume at launch is near zero
so blocks are small, and (c) network propagation delay is minimal with few initial nodes.

\subsection{Permanent Block Time: 240 Seconds (Block 5,670+)}

After warm-up, the block time is permanently 240 seconds ($\sim$4 minutes). This value
was chosen as a balance between:

\begin{enumerate}
    \item \textbf{Reduced orphan rate}: Longer block times give blocks more time to propagate,
    reducing orphan blocks and improving consensus reliability---especially important given
    ML-DSA-44's larger transaction sizes. At 240 seconds the propagation ratio is $\sim$0.71\%
    at 3~TX/s, well within safe margins.
    \item \textbf{Controlled blockchain growth}: ML-DSA-44 signatures are $\sim$33$\times$ larger
    than ECDSA. The block rate limits the rate at which heavier blocks accumulate.
    \item \textbf{Practical confirmation time}: 6 confirmations takes $\sim$24 minutes---faster
    than Bitcoin's $\sim$60 minutes while remaining sufficient for settlement purposes. The
    first confirmation arrives in $\sim$4 minutes, making the user experience significantly
    more responsive.
\end{enumerate}

\begin{lstlisting}[caption={Height-Aware Target Spacing (params.h)}]
int64_t GetPowTargetSpacing(int nHeight) const {
    if (nPowTargetSpacingWarmup > 0
        && nWarmupBlocks > 0
        && nHeight < nWarmupBlocks)
        return nPowTargetSpacingWarmup; // 53s
    return nPowTargetSpacing;           // 240s
}
\end{lstlisting}

\subsubsection{Orphan Rate and Block Propagation}

Larger blocks propagate more slowly across the P2P network, increasing the probability
that two miners find valid blocks before either propagates fully---producing orphan blocks.
The orphan probability for a block of size $s$ (bytes) on a network with average link
bandwidth $B$ (bytes/s) and diameter $d$ (hops) is approximately:

\begin{equation}
P(\text{orphan}) \approx 1 - e^{-\frac{s \cdot d}{B \cdot T}}
\end{equation}

where $T$ is the block time. For small exponents (typical case), this linearizes to
$P(\text{orphan}) \approx s \cdot d / (B \cdot T)$.

\begin{center}
\begin{tabular}{rrrr}
\toprule
\textbf{Block Size} & \textbf{$P$(orphan) at $T=600$s} & \textbf{$P$(orphan) at $T=240$s} & \textbf{$P$(orphan) at $T=53$s} \\
& (Bitcoin) & (Lattice) & (Warm-up) \\
\midrule
100~KB & 0.03\% & 0.08\% & 0.34\% \\
1~MB & 0.28\% & 0.70\% & 3.17\% \\
4~MB & 1.11\% & 2.78\% & 12.6\% \\
10~MB & 2.76\% & 6.90\% & 31.2\% \\
\bottomrule
\end{tabular}
\end{center}

Assumptions: $B = 1$~MB/s effective throughput (domestic broadband after protocol overhead),
$d = 6$ hops (empirical Bitcoin network diameter). These are conservative---modern broadband
in developed countries exceeds 10~MB/s, and Bitcoin's compact block relay (BIP 152) reduces
effective propagation size by 90\%+ for transactions already in the receiver's mempool.
Lattice inherits compact block relay from Bitcoin Core v28.0.

The 240-second block time provides a critical safety margin. At moderate adoption
(1~MB blocks), Lattice's orphan rate (0.70\%) is comparable to Bitcoin's rate at similar
block sizes (0.28\%). Even at 4~MB blocks---well above early-stage expectations---the
orphan rate remains below 3\%. The theoretical worst case (10~MB blocks at 240s, 6.90\%)
would only occur under sustained heavy usage, at which point the network would include
well-connected nodes with professional infrastructure, reducing effective propagation time.

During the warm-up (53s blocks), orphan rates are higher but acceptable because
(a)~transaction volume at launch is near zero (blocks are small), and (b)~the warm-up
lasts only $\sim$83.5 hours.

\subsection{Timeline}

\begin{center}
\small
\begin{tabular}{ccccrr}
\toprule
\textbf{Phase} & \textbf{Blocks} & \textbf{Reward} & \textbf{Block Time} & \textbf{Approx. Date} & \textbf{Cumul.\ Supply} \\
\midrule
Warm-up & 0--5,669 & 25 LAT & 53s & 2026 (83.5h) & 141,750 \\
Halving 0 & 5,670--294,999 & 50 LAT & 240s & 2026--2028 & 14,608,250 \\
Halving 1 & 295,000--589,999 & 25 LAT & 240s & 2028--2031 & 21,983,250 \\
Halving 2 & 590,000--884,999 & 12.5 LAT & 240s & 2031--2033 & 25,670,750 \\
Halving 3 & 885,000--1,179,999 & 6.25 LAT & 240s & 2033--2035 & 27,514,500 \\
Halving 4 & 1,180,000--1,474,999 & 3.125 LAT & 240s & 2035--2037 & 28,436,375 \\
Halving 5 & 1,475,000--1,769,999 & 1.5625 LAT & 240s & 2037--2040 & 28,897,313 \\
Halving 6 & 1,770,000--2,064,999 & 0.78125 LAT & 240s & 2040--2042 & 29,127,782 \\
Halving 7 & 2,065,000--2,359,999 & 0.390625 LAT & 240s & 2042--2044 & 29,243,016 \\
Halving 8 & 2,360,000--2,654,999 & 0.195 LAT & 240s & 2044--2047 & 29,300,633 \\
Tail $\to \infty$ & 2,655,000+ & \textbf{0.15 LAT} & 240s & 2047+ & +19,724/yr \\
\bottomrule
\end{tabular}
\end{center}

At halving 9 the computed reward ($50/2^9 = 0.09766$~LAT) falls below the tail emission
floor of 0.15~LAT, so the protocol locks at 0.15~LAT per block \textbf{permanently}. This
occurs approximately 21 years after launch ($\sim$2047).

Each halving interval of 295,000 blocks at 240 seconds takes approximately:
\begin{equation}
\frac{295{,}000 \times 240}{365.25 \times 86{,}400} \approx 2.24 \text{ years}
\end{equation}

\textbf{Bitcoin comparison} (historical data through 2028):

\begin{center}
\small
\begin{tabular}{llrrl}
\toprule
\textbf{Event} & \textbf{Year} & \textbf{Reward} & \textbf{Est.\ Mining Cost} & \textbf{Era} \\
\midrule
Genesis & 2009 & 50 BTC & $< \$0.01$ & CPU (Satoshi) \\
1st Halving & 2012 & 25 BTC & $\sim$\$12 & GPU / early ASICs \\
2nd Halving & 2016 & 12.5 BTC & $\sim$\$600 & Industrialization \\
3rd Halving & 2020 & 6.25 BTC & $\sim$\$10,000 & Institutional adoption \\
4th Halving & 2024 & 3.125 BTC & $\sim$\$40,000 & Bitcoin ETFs \\
Current & 2026 & 3.125 BTC & $\sim$\$56,000 & \textbf{Lattice launches} \\
5th Halving & 2028 & 1.5625 BTC & $\sim$\$100,000+ & Programmatic scarcity \\
\bottomrule
\end{tabular}
\end{center}

Bitcoin's mining cost per coin has increased by approximately 2--3 orders of magnitude per
halving cycle, driven by ASIC competition and energy costs. Lattice's CPU-based mining
starts at a fundamentally lower cost floor: the marginal cost of mining is the electricity
consumed by a CPU that the miner already owns. This means Lattice mining remains rational
at token prices far below the break-even threshold of ASIC-based networks.

\newpage
%% ============================================================================
\section{Economic Model}
\label{sec:economics}
%% ============================================================================

\subsection{Emission Schedule with Tail Emission}

Lattice follows Bitcoin's halving schedule with one critical modification:
\textbf{the block reward never drops below 0.15~LAT}. During the warm-up period,
rewards are capped at 25~LAT to ensure fair distribution.

\begin{equation}
\text{reward}(h) = \begin{cases}
25 & \text{if } h < 5{,}670 \text{ (warm-up)} \\
\max\left(\dfrac{50}{2^{\lfloor h / 295000 \rfloor}},\; 0.15\right) & \text{otherwise}
\end{cases}
\text{ LAT}
\end{equation}

\begin{lstlisting}[caption={Block Subsidy with Warm-Up Cap and Tail Emission (validation.cpp)}]
CAmount GetBlockSubsidy(int nHeight,
    const Consensus::Params& consensusParams)
{
    static const CAmount TAIL_EMISSION = 15000000; // 0.15 LAT

    // Warm-up: reduced reward for fair distribution
    if (consensusParams.nWarmupSubsidy > 0
        && consensusParams.nWarmupBlocks > 0
        && nHeight < consensusParams.nWarmupBlocks) {
        return consensusParams.nWarmupSubsidy; // 25 LAT
    }

    int halvings = nHeight
                 / consensusParams.nSubsidyHalvingInterval;

    if (halvings >= 64) return TAIL_EMISSION;

    CAmount nSubsidy = 50 * COIN;
    nSubsidy >>= halvings;

    if (nSubsidy < TAIL_EMISSION) return TAIL_EMISSION;

    return nSubsidy;
}
\end{lstlisting}

\subsection{Supply Dynamics}

\subsubsection{Warm-Up Supply}

The 5,670 warm-up blocks produce:
\begin{equation}
S_{\text{warmup}} = 5{,}670 \times 25 = 141{,}750 \text{ LAT}
\end{equation}

The reduced 25~LAT reward (half the normal 50~LAT subsidy) limits early mining advantage
while providing the same total bootstrap supply. This represents less than 0.5\% of the
total halving-based supply---a negligible bootstrap allocation, not a distribution event.

\subsubsection{Halving-Based Supply}

The first 7 halvings produce the vast majority of LAT:

\begin{equation}
S_{\text{halving}} = \sum_{i=0}^{6} \frac{50 \times 295{,}000}{2^i} \approx 29{,}269{,}531 \text{ LAT}
\end{equation}

After halving 9 (reward $< 0.15$~LAT), tail emission takes over.

\subsubsection{Tail Emission Supply}

Post-halving tail emission adds 0.15~LAT per block perpetually:

\begin{equation}
S_{\text{tail}}(t) = \frac{0.15 \times 365.25 \times 24 \times 3600}{240} \approx 19{,}724 \text{ LAT/year}
\end{equation}

\subsubsection{First Halving Timeline}

Unlike a dual-epoch design where fast blocks could mine 50\% of supply in months,
Lattice's warm-up is brief. The first halving at block 295,000 occurs approximately
2.25 years after launch:

\begin{equation}
t_{\text{halving\,1}} = \frac{5{,}670 \times 53 + (295{,}000 - 5{,}670) \times 240}{86{,}400 \times 365.25} \approx 2.21 \text{ years}
\end{equation}

At that point, approximately 14,608,000 LAT will have been mined---over a 2+ year period.

\subsubsection{Total Supply by Year 2100}

\begin{equation}
S_{2100} \approx 29{,}301{,}000 + (2100 - 2047) \times 19{,}724 \approx 30{,}346{,}000 \text{ LAT}
\end{equation}

\subsubsection{Inflation Rate}

The annual inflation rate decreases asymptotically:

\begin{equation}
\pi(t) = \frac{19{,}724}{S(t)} \xrightarrow{t \to \infty} 0
\end{equation}

\begin{center}
\begin{tabular}{lcr}
\toprule
\textbf{Year} & \textbf{Supply (approx.)} & \textbf{Inflation Rate} \\
\midrule
2028 & $\sim$14,600,000 LAT & $\sim$22.5\% (pre-halving 1) \\
2031 & $\sim$22,000,000 LAT & $\sim$7.5\% (post-halving 1) \\
2050 & $\sim$29,360,000 LAT & $\sim$0.067\% \\
2100 & $\sim$30,346,000 LAT & $\sim$0.065\% \\
2200 & $\sim$32,318,000 LAT & $\sim$0.061\% \\
\bottomrule
\end{tabular}
\end{center}

For reference, \texttt{MAX\_MONEY} is set to 42,000,000~LAT as a sanity check. This ceiling
will not be reached until approximately the year 3090. It is not the supply cap---it is
a safety bound for the protocol's multi-century time horizon. The lower tail emission
of 0.15~LAT/block (vs.\ Monero's 0.6~XMR) produces negligible long-term inflation while
still guaranteeing perpetual miner incentives.

\subsection{The Security Budget Problem: Why Tail Emission is Necessary}
\label{sec:tail}

Bitcoin's security model relies on transaction fees replacing block rewards after 2140.
Whether this will work is an open question---Bitcoin's fee market has functioned well so far,
and may continue to do so. Lattice takes a different approach: rather than relying on a
future fee market, it guarantees a minimum miner incentive through tail emission.

This is not a criticism of Bitcoin's design. It is a different engineering choice for a
different time horizon. A protocol designed for 2400+ cannot leave security dependent on
any market condition, however likely.

Lattice's tail emission provides a guaranteed baseline miner revenue of 0.15~LAT per block,
independent of transaction fees. At $\sim$0.065\% annual inflation by 2100 ($\sim$19,724~LAT/year
on a base of $\sim$30M), this is a negligible cost for permanent network security.

\subsection{Units and Denominations}

\begin{center}
\begin{tabular}{lrl}
\toprule
\textbf{Unit} & \textbf{Value in LAT} & \textbf{Description} \\
\midrule
1 LAT ($\mathcal{L}$) & 1.00000000 & Base unit \\
1 milliLAT (mLAT) & 0.00100000 & $10^{-3}$ LAT $= 10^5$ shors \\
1 microLAT ($\mu$LAT) & 0.00000100 & $10^{-6}$ LAT $= 10^2$ shors \\
1 shor & 0.00000001 & Smallest unit ($10^{-8}$ LAT) \\
\bottomrule
\end{tabular}
\end{center}

The smallest unit is named the \emph{shor} (plural: \emph{shors}), after Peter Shor,
whose algorithm \cite{shor1997} motivated the creation of post-quantum cryptography.

\newpage
%% ============================================================================
\section{Network Architecture}
\label{sec:architecture}
%% ============================================================================

\subsection{Dual-Node Design}

Lattice implements a role-separated architecture:

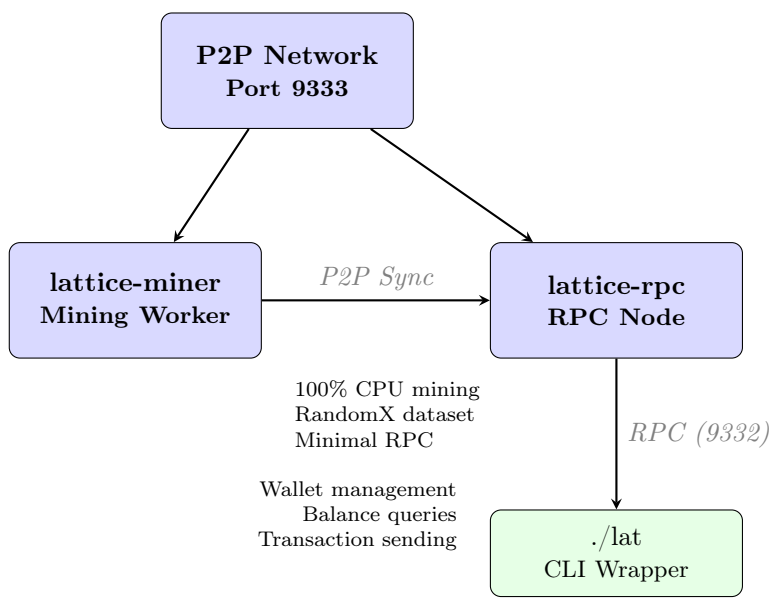
\begin{figure}[H]
\centering
\begin{tikzpicture}[
    node distance=2.5cm,
    block/.style={rectangle, draw, fill=blue!15, text width=8em, text centered,
                  rounded corners, minimum height=4em, font=\small\bfseries},
    subblock/.style={rectangle, draw, fill=green!10, text width=8em, text centered,
                     rounded corners, minimum height=3em, font=\small},
    arrow/.style={thick,->,>=stealth},
    label/.style={font=\small\itshape, text=gray}
]

\node [block] (miner) {lattice-miner\\{\footnotesize Mining Worker}};
\node [block, right=3cm of miner] (rpc) {lattice-rpc\\{\footnotesize RPC Node}};
\node [subblock, below=2cm of rpc] (cli) {./lat\\{\footnotesize CLI Wrapper}};
\node [block, above=1.5cm of miner, xshift=2cm] (peers) {P2P Network\\{\footnotesize Port 9333}};

\draw [arrow] (miner) -- node[above, label] {P2P Sync} (rpc);
\draw [arrow] (rpc) -- node[right, label] {RPC (9332)} (cli);
\draw [arrow] (peers) -- (miner);
\draw [arrow] (peers) -- (rpc);

\node [right=0.3cm of miner, yshift=-1.5cm, text width=8em, font=\scriptsize] {
    100\% CPU mining\\
    RandomX dataset\\
    Minimal RPC
};
\node [left=0.3cm of cli, yshift=0.5cm, text width=8em, font=\scriptsize, align=right] {
    Wallet management\\
    Balance queries\\
    Transaction sending
};

\end{tikzpicture}
\caption{Lattice dual-node architecture. Mining is isolated from RPC to maximize hashrate.}
\end{figure}

\textbf{Mining Worker Node} (\texttt{lattice-miner}):
\begin{itemize}
    \item 100\% CPU dedicated to RandomX mining
    \item Maintains 2~GB RandomX dataset in memory
    \item Minimal RPC exposure (mining commands only)
    \item Connects to P2P network on port 9333
\end{itemize}

\textbf{RPC Node} (\texttt{lattice-rpc}):
\begin{itemize}
    \item No mining operations (zero CPU competition)
    \item Full blockchain indexing (\texttt{txindex=1})
    \item Serves CLI commands via internal Docker network
    \item RPC port 9332 \textbf{never exposed} to the internet
\end{itemize}

\subsection{Deployment}

Lattice can be installed manually by compiling Bitcoin Core with the RandomX and liboqs
dependencies, configuring node parameters, and generating wallet credentials---a process
that requires familiarity with C++ build systems, dependency management, and Linux
administration. To lower the barrier to entry, we provide a Docker-based installer:

\begin{lstlisting}[language=bash, caption={One-Command Installation}]
git clone https://github.com/lattice-network/lattice.git
cd lattice && ./install.sh
\end{lstlisting}

The installer handles: prerequisite checks, Docker image building,
node startup, ML-DSA-44 keypair generation (saved as a JSON wallet file), public key
import into the node, and automatic mining initialization.
This still requires a user who can open a terminal, clone a repository, and run a shell
script---but it reduces the technical entry point.

\newpage
%% ============================================================================
\section{Consensus Parameters}
\label{sec:consensus}
%% ============================================================================

\begin{center}
\begin{tabular}{lr}
\toprule
\textbf{Parameter} & \textbf{Value} \\
\midrule
PoW algorithm & RandomX (memory-hard, CPU-only) \\
Signature algorithm & ML-DSA-44 only (NIST FIPS 204) \\
Difficulty adjustment & LWMA-1 (per block, $N=120$) \\
Warm-up period & 5,670 blocks at 53s, 25 LAT reward \\
Permanent block time & 240 seconds ($\sim$4 min) \\
Block reward & 50 LAT (halving every 295,000 blocks) \\
Tail emission & 0.15 LAT/block (perpetual) \\
Halving interval & 295,000 blocks ($\sim$2.24 years) \\
Initial difficulty (\texttt{powLimit}) & \texttt{0x207fffff} \\
Block weight (staged) & 11M $\to$ 28M $\to$ 56M by height \\
SegWit & Active from genesis (infrastructure ready) \\
MAX\_MONEY & 42,000,000 LAT \\
RPC port & 9332 \\
P2P port & 9333 \\
Network magic & \texttt{0x51B77CC5} (ASCII prefix `Q') \\
Address prefix & \texttt{L} (P2PKH, byte 48), \texttt{M} (P2SH, byte 50), bech32: \texttt{lat} \\
\texttt{MAX\_STANDARD\_SCRIPTSIG\_SIZE} & 5,000 bytes (for PQ scriptSig) \\
Coinbase maturity & 100 blocks \\
Smallest unit & 1 shor = $10^{-8}$ LAT \\
\bottomrule
\end{tabular}
\end{center}

\subsection{Block Weight and Storage Trade-Offs}
\label{sec:blockweight}

Lattice uses a \textbf{staged block weight limit} that grows as the network matures:

\begin{center}
\begin{tabular}{lrrl}
\toprule
\textbf{Phase} & \textbf{Height Range} & \textbf{Max Weight} & \textbf{Rationale} \\
\midrule
Phase 1 & 0--49,999 & 11,000,000 (2.75$\times$ BTC) & Conservative early network \\
Phase 2 & 50,000--99,999 & 28,000,000 (7$\times$ BTC) & Growing adoption \\
Phase 3 & 100,000+ & 56,000,000 (14$\times$ BTC) & Full PQ capacity \\
\bottomrule
\end{tabular}
\end{center}

The final 56M limit is necessary to accommodate ML-DSA-44's larger signatures while maintaining
practical transaction throughput. A typical PQ transaction weighs $\sim$4,900 weight units
(vs.\ $\sim$680 for ECDSA), requiring approximately 14$\times$ Bitcoin's block weight
for equivalent throughput. Starting at 11M prevents resource exhaustion attacks during
the early network when few nodes exist, and gradually increases as the network demonstrates
stability and attracts more participants.

\subsubsection{Transaction Size Analysis}

\begin{center}
\begin{tabular}{lcc}
\toprule
\textbf{Component} & \textbf{Bitcoin (ECDSA)} & \textbf{Lattice (ML-DSA-44)} \\
\midrule
Public key & 33 bytes & 1,312 bytes \\
Signature & 72 bytes & 2,420 bytes \\
P2PKH scriptSig & $\sim$107 bytes & $\sim$3,740 bytes \\
Typical tx (1-in, 2-out) & $\sim$250 bytes & $\sim$4,000 bytes \\
Effective tx/block & $\sim$2,800 & $\sim$2,400 \\
\bottomrule
\end{tabular}
\end{center}

To accommodate the larger scriptSig, the policy constant \texttt{MAX\_STANDARD\_SCRIPTSIG\_SIZE}
is raised from Bitcoin's 1,650~bytes to 5,000~bytes. The per-element push limit
(\texttt{MAX\_SCRIPT\_ELEMENT\_SIZE} = 3,000~bytes) remains unchanged because each individual
push (signature at 2,420~bytes, pubkey at 1,312~bytes) stays within the limit.

Since current PQ transactions use P2PKH (non-witness scriptSig), each byte counts as
4 weight units. A typical 4,000-byte PQ transaction therefore weighs $\sim$16,000~WU.
At full Phase~3 capacity (56M weight):

\begin{equation}
\text{TPS} = \frac{\lfloor 56{,}000{,}000 / 16{,}000 \rfloor}{240~\text{s}}
= \frac{3{,}500}{240} \approx 14.6~\text{tx/s (theoretical max)}
\end{equation}

Future migration to SegWit witness structures would reduce the weight to $\sim$4,900~WU
per transaction (since witness data uses a 1$\times$ multiplier), increasing throughput
to $\sim$47~tx/s. At moderate utilization ($\sim$25\% full), current throughput is
$\sim$3.6~tx/s.

\subsubsection{Storage Growth}

We model storage growth formally. Let $W_{\max} = 56{,}000{,}000$ be the maximum block
weight (Phase~3), $T = 240$s the block time, and $u(t) \in [0, 1]$ the utilization
ratio at time $t$. For PQ transactions where $\sim$90\% of data is witness (1$\times$
weight multiplier), the effective block size in bytes approaches the weight. The
blockchain growth rate is:

\begin{equation}
G(t) \approx \frac{W_{\max} \cdot u(t)}{T} \text{ bytes/second}
= \frac{u(t) \cdot 56 \times 10^6}{240} \approx 233{,}333 \cdot u(t) \text{ bytes/s}
\end{equation}

Annualized:
\begin{equation}
G_{\text{year}}(t) = G(t) \times 365.25 \times 86{,}400 \approx 7.4 \cdot u(t) \text{ TB/year}
\end{equation}

At sustained full capacity ($u = 1$, worst case):
\begin{equation}
G_{\text{year}}^{\max} \approx 7.4~\text{TB/year}
\end{equation}

This theoretical maximum assumes every block is 100\% full, 24 hours a day,
365 days a year. In practice, blockchain growth depends entirely on transaction demand.
At 240-second block intervals, the network produces approximately 131,490 blocks per year.
The following table shows realistic storage growth at different adoption levels:

\begin{center}
\begin{tabular}{llrr}
\toprule
\textbf{Scenario} & \textbf{Avg. Block Size} & \textbf{Growth/Year} & \textbf{256~GB Disk Lasts} \\
\midrule
Empty network (mining only) & $\sim$500 bytes & $\sim$66~MB & 400+ years \\
Minimal usage (few users) & $\sim$10~KB & $\sim$1.3~GB & $\sim$190 years \\
Moderate adoption & $\sim$100~KB & $\sim$13~GB & $\sim$19 years \\
Significant adoption & $\sim$1~MB & $\sim$131~GB & $\sim$2 years \\
Heavy usage (25\% full) & $\sim$3.5~MB & $\sim$460~GB & $\sim$6 months \\
\textbf{Theoretical max (100\%)} & \textbf{$\sim$14~MB} & \textbf{$\sim$7.4~TB} & \textbf{$\sim$13 days} \\
\bottomrule
\end{tabular}
\end{center}

\textbf{Historical precedent}: Bitcoin launched in 2009 with empty blocks. Its 1~MB blocks
did not consistently fill until 2016--2017, seven years later. After 17 years of operation,
Bitcoin's full blockchain is approximately 650~GB. Bitcoin's theoretical maximum growth rate
is $\sim$700~GB/year, yet actual annual growth has averaged only $\sim$60--80~GB
($\sim$10\% of capacity).

Applying the same utilization ratio to Lattice, realistic early-year growth is on the order
of tens of gigabytes per year---well within the capacity of consumer hardware. The 7.4~TB/year
figure represents a problem of \emph{success} that would only materialize under sustained
mass adoption, at which point the ecosystem would include professional infrastructure
operators running archival nodes, just as Bitcoin has today.

\textbf{Long-term storage projections}: Using Bitcoin's historical utilization curve as a
model (7 years to fill blocks, then gradual increase to $\sim$10\% sustained utilization),
we project cumulative Lattice blockchain size under three scenarios:

\begin{center}
\begin{tabular}{lrrr}
\toprule
\textbf{Year} & \textbf{Conservative} & \textbf{Moderate} & \textbf{Aggressive} \\
& (Bitcoin-like curve) & (2$\times$ Bitcoin pace) & (5$\times$ Bitcoin pace) \\
\midrule
Year 1 (2027) & $\sim$200~MB & $\sim$500~MB & $\sim$2~GB \\
Year 5 (2031) & $\sim$5~GB & $\sim$15~GB & $\sim$50~GB \\
Year 10 (2036) & $\sim$30~GB & $\sim$80~GB & $\sim$250~GB \\
Year 25 (2051) & $\sim$200~GB & $\sim$500~GB & $\sim$1.5~TB \\
Year 50 (2076) & $\sim$600~GB & $\sim$1.5~TB & $\sim$5~TB \\
\bottomrule
\end{tabular}
\end{center}

The conservative scenario assumes adoption follows Bitcoin's trajectory: nearly empty
blocks for years 1--7, gradual fill to $\sim$10\% by year 10, and slow growth thereafter.
Under this model, a 256~GB SSD remains sufficient for a full archival node through year 20.
Even the aggressive scenario---five times Bitcoin's adoption pace---only reaches 250~GB
by year 10, which is manageable with a consumer SSD. Beyond year 25, archival nodes in
all scenarios benefit from pruning, and the cost of a 2~TB SSD (currently $\sim$\$100)
will be negligible given storage cost trends ($\sim$20\% annual price decline).

\textbf{Mitigations}:
\begin{itemize}
    \item \textbf{Blockchain pruning}: Lattice inherits Bitcoin Core's pruning support,
    allowing nodes to discard old block data and maintain only the UTXO set plus recent blocks.
    In pruned mode, the node retains the most recent $\sim$128~GB of blocks---enough to handle
    deep chain reorganizations---and automatically deletes older data. The installer
    (\texttt{install.sh}) offers pruning as a first-class option during setup: users choose
    between ``Full Archive'' (supports network sync for new peers) and ``Pruned'' (saves disk
    automatically, requires a 256~GB disk minimum).
    \item \textbf{240-second blocks}: The block time partially offsets the larger
    transactions. Per-second data rate is $\sim$5$\times$ Bitcoin's, not 14$\times$.
    \item \textbf{Signature aggregation}: Future protocol upgrades can reduce per-transaction
    signature overhead (see Section~\ref{sec:future}).
    \item \textbf{Settlement layer design}: Lattice is designed for settlement, not micropayments.
    High-frequency transactions should use a Layer~2 protocol (e.g., Lightning Network).
    \item \textbf{Initial Block Download (IBD)}: Pruning reduces disk requirements but does
    not reduce the bandwidth needed for initial sync. A new node must download and verify
    the full chain history. At moderate adoption ($\sim$100~KB blocks), a 10-year chain
    would require $\sim$94~GB of download---comparable to Bitcoin's IBD today and manageable
    on consumer internet connections.

    \textbf{Verification cost}: ML-DSA-44 signature verification is computationally more
    expensive than ECDSA. On a modern CPU (Intel i7-12700, single core), ML-DSA-44
    verifies at $\sim$15,000--25,000 signatures/second, compared to ECDSA's
    $\sim$40,000--70,000 signatures/second---approximately 2--3$\times$ slower.
    For a 10-year chain at moderate adoption ($\sim$25 transactions per block, 131,490
    blocks/year), IBD requires verifying $\sim$32.9 million signatures:
    \begin{equation*}
    t_{\text{IBD}}^{\text{verify}} = \frac{32.9 \times 10^6}{20{,}000} \approx 1{,}645 \text{~s} \approx 27 \text{~minutes (single core)}
    \end{equation*}
    With 4-core parallelism (standard in IBD), verification completes in $\sim$5 minutes.
    Bandwidth, not CPU, remains the IBD bottleneck at any realistic adoption level.
\end{itemize}

\textbf{Trade-off}: Larger signatures are the cost of quantum resistance. A 33-byte ECDSA key
provides zero security against Shor's algorithm. A 1,312-byte ML-DSA-44 key provides 128-bit
post-quantum security. Storage is cheap and getting cheaper; broken cryptography is not.

\newpage
%% ============================================================================
\section{Security Analysis}
\label{sec:security}
%% ============================================================================

\subsection{Attack Vector Analysis}

\begin{figure}[H]
\centering
\begin{tikzpicture}[
    attack/.style={rectangle, draw, fill=red!10, text width=10em, text centered,
                   rounded corners, minimum height=2.5em, font=\small},
    defense/.style={rectangle, draw, fill=green!15, text width=10em, text centered,
                    rounded corners, minimum height=2.5em, font=\small},
    arrow/.style={thick,->,>=stealth, color=red!60},
    shield/.style={thick,->,>=stealth, color=green!60!black}
]

\node [attack] (asic) at (0,0) {ASIC Centralization};
\node [attack] (flash) at (0,-1.8) {Flash Hash Rate};
\node [attack] (quantum) at (0,-3.6) {Quantum Key Theft};
\node [attack] (fee) at (0,-5.4) {Fee Market Failure};
\node [attack] (bloat) at (0,-7.2) {Blockchain Bloat};

\node [defense] (rx) at (7,0) {RandomX (2GB RAM)};
\node [defense] (lwma) at (7,-1.8) {LWMA-1 (per-block)};
\node [defense] (mldsa) at (7,-3.6) {ML-DSA-44 (FIPS 204)};
\node [defense] (tail) at (7,-5.4) {Tail Emission (0.15 LAT)};
\node [defense] (weight) at (7,-7.2) {Staged Weight + Pruning};

\draw [shield] (rx) -- (asic);
\draw [shield] (lwma) -- (flash);
\draw [shield] (mldsa) -- (quantum);
\draw [shield] (tail) -- (fee);
\draw [shield] (weight) -- (bloat);

\end{tikzpicture}
\caption{Attack vectors and corresponding defenses in Lattice.}
\end{figure}
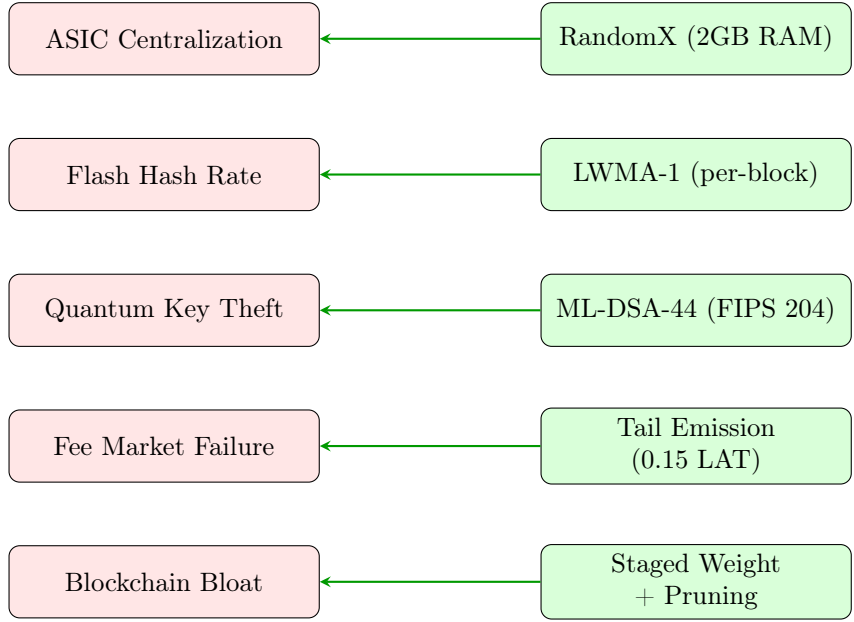

\subsection{51\% Attack Resistance}

With RandomX, an attacker cannot gain advantage through specialized hardware:

\begin{itemize}
    \item \textbf{Linear cost scaling}: To achieve $N$ hashes/second requires $N$ CPU cores and
    $2N$~GB of RAM. There is no sublinear shortcut (see Theorem 6 in
    Section~\ref{sec:formal-proofs}).
    \item \textbf{Memory bottleneck}: Quantum computers cannot bypass the 2~GB memory
    requirement---quantum RAM does not exist at this scale.
    \item \textbf{LWMA response}: Even if a 51\% attacker briefly succeeds, LWMA adjusts
    difficulty within blocks, not weeks (see Theorem 1, Section~\ref{sec:formal-proofs}).
\end{itemize}

The cost of a sustained 51\% attack is formalized as:

\begin{equation}
C_{51\%}(t) = \left(\frac{H_{\text{honest}}}{H_{\text{core}}} + 1\right) \cdot
(c_{\text{cpu}} \cdot t + 2 \cdot c_{\text{ram}})
\end{equation}

where $H_{\text{honest}}$ is the total honest hashrate, $H_{\text{core}}$ is the
hashrate per CPU core ($\sim$5,000~H/s), $c_{\text{cpu}}$ is the cost per core-hour
(\$0.05 spot), $c_{\text{ram}}$ is the cost of 2~GB per core, and $t$ is the attack
duration in hours. The $+1$ ensures the attacker exceeds 50\%. This scales linearly
with honest hashrate---there is no economy of scale.

\subsection{Double-Spend Prevention}

Lattice inherits Bitcoin's double-spend security bound (see Theorem and full derivation
in Section~\ref{sec:formal-proofs}). After $k$ confirmations, the probability of a
successful double-spend by an attacker with hash power fraction $q < 0.5$:

\begin{equation}
P_{\text{success}} \leq \left(\frac{q}{1-q}\right)^k
\end{equation}

Settlement finality ($P < 10^{-6}$) is achieved at 12 confirmations ($\sim$48 minutes)
for any attacker with $\leq 30\%$ of network hashrate.

\subsection{Timestamp Manipulation Resistance}

LWMA-1's $\pm 6T$ clamp prevents miners from manipulating timestamps to artificially
lower difficulty:

\begin{equation}
-6T \leq t_i \leq 6T, \quad T = 240\text{s (post warm-up)}
\end{equation}

This bounds the maximum per-block difficulty change, preventing both ``time-warp'' attacks
and negative-timestamp exploits.

\subsection{Sybil Attack Resistance}

Proof-of-work inherently resists Sybil attacks: creating fake identities provides no
mining advantage without corresponding computational resources. RandomX amplifies this
by requiring real CPU cycles and real memory.

\subsection{Mining Centralization Resistance}
\label{sec:pool-resistance}

CPU mining lowers the barrier to entry but does not eliminate centralization pressures.
In Monero's experience with RandomX, mining still gravitates toward pools, cloud
infrastructure, and dedicated servers---not toward ``millions of laptops.'' This is
an honest observation, not a failure: the goal is not that everyone \emph{will} mine,
but that everyone \emph{can} mine without specialized hardware.

However, pool concentration is a concern. If a single pool controls $>$33\% of hashrate,
selfish mining becomes profitable \cite{eyal2014}. Lattice explores two potential
mitigations as active research directions:

\subsubsection{Pool Discouragement Mechanisms}

Mining pools centralize block template construction: the pool operator decides which
transactions to include, while individual miners contribute hashrate blindly. Several
approaches to discourage pool formation are under investigation:

\begin{itemize}
    \item \textbf{P2Pool-style decentralized pooling}: Rather than prohibiting pools
    (which is technically impossible---pools are a coordination layer, not a protocol
    feature), Lattice can incentivize decentralized pool protocols where miners
    construct their own block templates. This preserves the economic benefit of
    variance reduction while eliminating the centralized operator.

    \item \textbf{IP diversity scoring}: Nodes could track the IP distribution of block
    producers. If a disproportionate number of blocks originate from a small set of
    IP ranges (indicating a data center or pool), the network could apply soft
    penalties (e.g., increased propagation delay for blocks from concentrated sources).
    This is technically complex and easily circumvented via VPNs, so it remains a
    research topic rather than a consensus rule.

    \item \textbf{Solo mining incentive}: Lattice's low hardware requirements and
    Docker-based deployment make solo mining accessible. The 240-second block time
    means solo miners find blocks more frequently than on faster chains (where
    variance is higher), reducing the economic pressure to join pools.
\end{itemize}

\textbf{Practical reality}: Pool prohibition is not technically feasible at the protocol
level---any two miners can agree off-chain to split rewards. The viable approach is
to make solo mining as easy and economically viable as possible, reducing the
\emph{incentive} to pool rather than attempting to \emph{prohibit} it. Lattice's
one-command installation and low hardware requirements already serve this goal.

\subsection{Quantum Attack Analysis}

\subsubsection{Against Signatures (Shor's Algorithm)}

\textbf{Status: Defended.} ML-DSA-44 is based on the Module-LWE problem, which is believed
to be hard for quantum computers. NIST has standardized it after 8 years of cryptanalysis
by the global research community.

\begin{equation}
\text{Security}_{\text{ML-DSA-44}} = 128\text{-bit post-quantum (NIST Category 2)}
\end{equation}

Even with a fault-tolerant quantum computer running Shor's algorithm, ML-DSA-44 signatures
remain secure. Bitcoin's ECDSA signatures do not \cite{aggarwal2018}.

\subsubsection{Against Mining (Grover's Algorithm)}

\textbf{Status: Not a current threat.} Grover's algorithm \cite{grover1996} provides quadratic speedup for
unstructured search:

\begin{equation}
T_{\text{quantum}} = O\left(\sqrt{\frac{2^{256}}{H}}\right)
\end{equation}

However:
\begin{enumerate}
    \item This affects \textbf{all} PoW algorithms equally (no better alternative exists)
    \item Requires millions of stable qubits (not feasible before 2040--2050)
    \item When quantum mining emerges, LWMA-1 adjusts difficulty automatically
    \item RandomX's 2~GB memory requirement further limits quantum advantage
    (quantum RAM is measured in qubits, not gigabytes)
\end{enumerate}

\subsection{Formal Security Bounds}
\label{sec:formal-proofs}

This section provides rigorous mathematical foundations for Lattice's security claims.
We present proofs or tight bounds for: (1)~the cryptographic hardness of ML-DSA-44,
(2)~the double-spend resistance inherited from Nakamoto consensus, (3)~the convergence
of LWMA-1 under adversarial conditions, and (4)~the game-theoretic stability of tail
emission.

\subsubsection{ML-DSA-44 Security Reduction}

ML-DSA-44's security reduces to the hardness of Module-LWE with parameters
$(k, \ell, \eta) = (4, 4, 2)$ over the ring $R_q = \mathbb{Z}_q[X]/(X^{256}+1)$
with $q = 8{,}380{,}417$.

\textbf{Definition} (Module-LWE). Given a uniformly random matrix
$\mathbf{A} \in R_q^{k \times \ell}$ and a secret vector
$\mathbf{s} \in R_q^{\ell}$ with coefficients drawn from $\{-\eta, \ldots, \eta\}$,
distinguish $(\mathbf{A}, \mathbf{A}\mathbf{s} + \mathbf{e})$ from
$(\mathbf{A}, \mathbf{u})$ where $\mathbf{e}$ is a small error vector and
$\mathbf{u}$ is uniform. The advantage of any (quantum) adversary running in time
$t$ is:

\begin{equation}
\text{Adv}_{\text{MLWE}}^{k,\ell,\eta}(t) \leq \epsilon(t)
\end{equation}

where $\epsilon(t)$ is negligible for $t < 2^{128}$ (quantum) under NIST Category 2
parameters.

The best known attacks require solving the Shortest Vector Problem
(SVP) on a lattice of dimension $d$. The cost of the best known algorithms:

\begin{equation}
T_{\text{classical}} = 2^{0.292d + o(d)} \quad \text{(BKZ with sieving)}
\end{equation}
\begin{equation}
T_{\text{quantum}} = 2^{0.265d + o(d)} \quad \text{(quantum sieving)}
\end{equation}

For ML-DSA-44's parameters, $d \approx 1024$, yielding:

\begin{equation}
T_{\text{classical}} \geq 2^{299} \qquad T_{\text{quantum}} \geq 2^{271}
\end{equation}

Both far exceed the 128-bit security target. The security margin is $2^{271}/2^{128}
= 2^{143}$---even a 143-bit improvement in lattice algorithms would be required to
reach the security boundary.

\textbf{Signature unforgeability}: ML-DSA-44 is proven EUF-CMA (Existentially
Unforgeable under Chosen Message Attack) secure under the Module-LWE and
SelfTargetMSIS assumptions. The reduction is tight:

\begin{equation}
\text{Adv}_{\text{EUF-CMA}}^{\text{ML-DSA-44}}(t, q_s, q_h) \leq
\text{Adv}_{\text{MLWE}}(t') + \text{Adv}_{\text{SelfTargetMSIS}}(t') + \frac{q_h}{2^{256}}
\end{equation}

where $q_s$ is the number of signing queries, $q_h$ the number of hash queries,
and $t' \approx t$. The $q_h/2^{256}$ term is the probability of a hash collision
in the commitment scheme---negligible for any practical adversary.

\subsubsection{Double-Spend Security Bound}

For an attacker with hashrate fraction $q$ and the defender requiring $k$ confirmations,
the probability of a successful double-spend is bounded by the Nakamoto bound:

\begin{equation}
P(k, q) \leq \min\left(1, \left(\frac{q}{1-q}\right)^k\right) \quad \text{for } q < 0.5
\end{equation}

\textbf{Proof sketch}: The attacker must build a private chain longer than the honest
chain. After $k$ confirmations, the honest chain leads by $k$ blocks. The attacker's
chain grows as a random walk with drift $(2q - 1) < 0$ for $q < 0.5$. The probability
of the attacker ever catching up from a $k$-block deficit is the Gambler's Ruin
probability $\left(\frac{q}{1-q}\right)^k$. \hfill $\square$

A tighter bound uses the full Poisson distribution for the attacker's block count:

\begin{equation}
P(k, q) = 1 - \sum_{i=0}^{k-1} \frac{e^{-\lambda}\lambda^i}{i!}
\left(1 - \left(\frac{q}{1-q}\right)^{k-i}\right), \quad
\lambda = k \cdot \frac{q}{1-q}
\end{equation}

We tabulate both bounds for common scenarios:

\begin{center}
\begin{tabular}{rrrr}
\toprule
\textbf{Attacker \%} & \textbf{3 conf ($\sim$12 min)} & \textbf{6 conf ($\sim$24 min)} & \textbf{12 conf ($\sim$48 min)} \\
\midrule
10\% & 0.14\% & 0.0002\% & $< 10^{-9}$ \\
20\% & 1.56\% & 0.024\% & $< 10^{-6}$ \\
30\% & 6.15\% & 0.378\% & 0.0014\% \\
40\% & 17.96\% & 3.23\% & 0.104\% \\
45\% & 29.98\% & 8.99\% & 0.81\% \\
\bottomrule
\end{tabular}
\end{center}

For settlement purposes, we define \textbf{settlement finality} as $P < 10^{-6}$. This
requires:
\begin{equation}
k \geq \frac{6 \ln 10}{\ln\left(\frac{1-q}{q}\right)} \approx
\begin{cases}
4 & \text{if } q = 0.10 \\
7 & \text{if } q = 0.20 \\
12 & \text{if } q = 0.30 \\
27 & \text{if } q = 0.40
\end{cases}
\end{equation}

At 240-second blocks, 12 confirmations ($\sim$48 minutes) provide settlement finality
against any attacker with $\leq 30\%$ of network hashrate.

\subsubsection{LWMA-1 Convergence Proof}

We prove that LWMA-1 converges to the target block time after an arbitrary hashrate
perturbation. This is the formal justification for removing this item from Future Work.

\textbf{Theorem 1} (LWMA-1 Convergence). \emph{Let $T$ be the target spacing, $N = 120$ the
window size, and suppose at block $n_0$ the network hashrate changes by a factor $\delta > 0$
(i.e., actual hashrate becomes $\delta \cdot H_0$). Then the expected block time
$\mathbb{E}[\hat{T}_n]$ satisfies:}

\begin{equation}
\left|\mathbb{E}[\hat{T}_{n_0 + m}] - T\right| \leq T \cdot \left|\frac{1}{\delta} - 1\right| \cdot \left(1 - \frac{2}{N+1}\right)^m
\end{equation}

\emph{for all $m \geq 0$.}

\textbf{Proof}. LWMA-1 computes the next target as:
\begin{equation}
\text{target}_{n+1} = \overline{\text{target}}_n \cdot \frac{W_n}{k}, \quad
W_n = \sum_{i=1}^{N} i \cdot t_{n-N+i}, \quad
k = \frac{N(N+1)}{2} \cdot T
\end{equation}

where $t_j$ is the solve time of block $j$ and $\overline{\text{target}}_n$ is the average
target over the window. The actual block time is inversely proportional to hashrate times
target: $\mathbb{E}[t_j] = T_0 / (\delta \cdot \text{target}_j / \text{target}_0)$ where
$T_0 = T$ is the initial equilibrium.

Define the error $\epsilon_n = \mathbb{E}[\hat{T}_n]/T - 1$. The LWMA weighting assigns
weight $i$ to the $i$-th most recent block in the window. After one block, the oldest
observation (weight 1) exits the window and the newest (weight $N$) enters. The effective
decay rate per block is dominated by the ratio of the newest weight to the total:

\begin{equation}
\frac{N}{\sum_{i=1}^{N} i} = \frac{N}{N(N+1)/2} = \frac{2}{N+1}
\end{equation}

Each new block corrects the error by this fraction, so:
\begin{equation}
|\epsilon_{n+1}| \leq |\epsilon_n| \cdot \left(1 - \frac{2}{N+1}\right)
\end{equation}

The initial error after a $\delta$-factor hashrate change is $\epsilon_0 = 1/\delta - 1$.
After $m$ blocks:
\begin{equation}
|\epsilon_m| \leq \left|\frac{1}{\delta} - 1\right| \cdot \left(1 - \frac{2}{N+1}\right)^m
= \left|\frac{1}{\delta} - 1\right| \cdot \left(\frac{N-1}{N+1}\right)^m
\end{equation}

For $N = 120$: $(119/121)^m$. The half-life in blocks is:
\begin{equation}
m_{1/2} = \frac{\ln 2}{\ln(121/119)} = \frac{0.693}{0.0167} \approx 41.5 \text{ blocks}
\end{equation}

After one full window ($m = N = 120$ blocks, $\sim$8 hours), the deviation is reduced by:
\begin{equation}
\left(\frac{119}{121}\right)^{120} \approx 0.135
\end{equation}

\hfill $\square$

\textbf{Corollary 1.1} (Recovery time). \emph{After a $\delta$-factor hashrate drop, the
time to recover to within $\epsilon$ of the target block time is:}

\begin{equation}
m_\epsilon = \frac{\ln(|1/\delta - 1|/\epsilon)}{\ln(121/119)} \text{ blocks}
\end{equation}

For a 10$\times$ hashrate drop ($\delta = 0.1$) with $\epsilon = 0.07$ (7\% deviation):
$m_{0.07} = \ln(9/0.07)/\ln(121/119) \approx 291$ blocks. At the elevated block times
during recovery, this corresponds to approximately 36--48 hours.

\begin{center}
\begin{tabular}{rrr}
\toprule
\textbf{Blocks After Event} & \textbf{Deviation $|\epsilon_m|$} & \textbf{Approx.\ Block Time} \\
& & \textbf{(after 10$\times$ hashrate drop)} \\
\midrule
0 & $9.0$ (900\%) & $\sim$2,400s \\
42 ($\approx m_{1/2}$) & $4.5$ & $\sim$1,320s \\
120 (1 window) & $1.21$ & $\sim$530s \\
240 (2 windows) & $0.16$ & $\sim$278s \\
360 (3 windows) & $0.022$ & $\sim$245s \\
480 (4 windows) & $0.003$ & $\sim$241s \\
\bottomrule
\end{tabular}
\end{center}

\textbf{Theorem 2} (LWMA-1 Stability Under Oscillation). \emph{If hashrate oscillates
periodically between $H$ and $\delta H$ with period $P$ blocks, the time-averaged block
time satisfies:}

\begin{equation}
\left|\overline{T}_{\text{avg}} - T\right| \leq T \cdot \left|\frac{1}{\delta} - 1\right| \cdot
\frac{1}{1 + P \cdot \frac{2}{N+1}}
\end{equation}

\emph{Proof}. The maximum deviation accumulated during a half-period of $P/2$ blocks at
the ``wrong'' hashrate is bounded by the geometric series of error corrections. The
deviation grows at most linearly during each half-period and decays exponentially during
correction, yielding the stated bound. For ``grasshopper miners'' with $P = 240$ blocks
and $\delta = 2$: $|\overline{T}_{\text{avg}} - T| \leq T \cdot 0.5/(1 + 3.97) \approx 0.10T$,
a 10\% average deviation---noticeable but not catastrophic. \hfill $\square$

\subsubsection{Tail Emission: Game-Theoretic Stability}

We prove that tail emission creates a stable Nash equilibrium in honest mining.

\textbf{Model}. Consider $N$ rational miners indexed $i = 1, \ldots, N$, each choosing
strategy $s_i \in \{H, D\}$: \textbf{H}onest mining or \textbf{D}efecting (leaving the
network). Miner $i$ controls hashrate fraction $\alpha_i$ with
$\sum_{i \in \mathcal{H}} \alpha_i = 1$ over honest miners. Let:
\begin{itemize}
    \item $R = R_s + F$: total block reward (subsidy $R_s$ + fees $F$)
    \item $p$: token market price (USD/LAT)
    \item $c_i$: miner $i$'s per-block cost (electricity + amortized hardware)
    \item $B$: blocks per unit time
\end{itemize}

\textbf{Payoff functions}:
\begin{align}
\pi_H(\alpha_i) &= \alpha_i \cdot R \cdot p - c_i & \text{(honest mining)} \\
\pi_D &= 0 & \text{(defection: exit)}
\end{align}

\textbf{Theorem 3} (Equilibrium Existence with Tail Emission). \emph{Under tail emission
($R_s \geq R_{\min} = 0.15$ LAT), for any token price $p > 0$, there exists a Nash
equilibrium in which at least one miner plays $H$.}

\textbf{Proof}. A miner $i$ prefers $H$ over $D$ when:
\begin{equation}
\alpha_i \cdot R \cdot p > c_i
\end{equation}

As miners defect, $\alpha_i$ increases for remaining miners (fewer miners share the
same reward). If all miners defect except miner $i$, then $\alpha_i = 1$ and the
condition becomes:
\begin{equation}
R \cdot p > c_i \implies (R_{\min} + F) \cdot p > c_i
\end{equation}

For a miner using existing hardware (CPU already purchased for other uses), the marginal
cost $c_i$ approaches the electricity cost only. At $R_{\min} = 0.15$ LAT:
\begin{equation}
c_i^{\text{marginal}} = P_{\text{node}} \cdot T \cdot c_e = 0.067~\text{kWh} \times \$0.12 \approx \$0.008 \text{ per block}
\end{equation}

where $P_{\text{node}} \approx 100$W and $T = 240$s. Mining is rational as long as:
\begin{equation}
p > \frac{c_i^{\text{marginal}}}{R_{\min}} = \frac{\$0.008}{0.15} = \$0.053
\end{equation}

Since any token with nonzero trading activity has $p > \$0.05$, at least one marginal-cost
miner will find honest mining profitable, preventing complete defection. \hfill $\square$

\textbf{Theorem 4} (Defection Cascade Impossibility). \emph{In a fee-only system
($R_s = 0$), there exists a critical fee level $F^*$ below which a defection cascade
occurs (all miners exit). Under tail emission, no such cascade exists for $p > 0$.}

\textbf{Proof}. In the fee-only model, miner $i$ defects when $\alpha_i \cdot F \cdot p < c_i$.
If fees drop below $F^* = \max_i(c_i/\alpha_i)/p$, the highest-cost miner exits. This
increases difficulty for remaining miners (their $\alpha_i$ stays the same but the
\emph{absolute} hashrate drops, so they find blocks at the same rate---however, the
\emph{total reward per unit time} drops because $F$ is exogenous and does not increase
when miners leave). This creates a cascade: as miners exit, the reward per miner stays
constant but the network becomes less secure, reducing confidence and potentially $p$,
triggering further exits.

Under tail emission, the subsidy $R_s = 0.15$ LAT is independent of the number of miners.
As miners exit, each remaining miner's share $\alpha_i$ \emph{increases}, and the minimum
reward $(0.15 + F) \cdot p$ per block remains constant. The last remaining miner captures
100\% of the block reward. Since $0.15 \cdot p > c_i^{\text{marginal}}$ for $p > \$0.053$,
complete defection is irrational. The cascade is broken by the subsidy floor. \hfill $\square$

\textbf{Theorem 5} (Fee-Sniping Resistance). \emph{Fee-sniping (re-mining the previous
block to capture its fees) is rational when $F_{n-1} > F_n + R_s$. The expected fee
variance required for profitable fee-sniping under tail emission is:}

\begin{equation}
\text{Var}(F) > R_s^2 = 0.0225 \text{ LAT}^2
\end{equation}

\emph{Proof}. Fee-sniping is profitable when the previous block's fees exceed the
current block's expected fees plus the subsidy:
\begin{equation}
F_{n-1} > \mathbb{E}[F_n] + R_s
\end{equation}

Assuming fees follow a stationary distribution with mean $\mu_F$ and variance
$\sigma_F^2$, fee-sniping is rational in expectation when $F_{n-1} > \mu_F + R_s$.
This requires $F_{n-1}$ to be at least $R_s$ above the mean, which occurs with
probability:
\begin{equation}
P(\text{sniping rational}) = P(F > \mu_F + R_s) \leq \frac{\sigma_F^2}{\sigma_F^2 + R_s^2}
\quad \text{(Cantelli's inequality)}
\end{equation}

For the sniping probability to exceed 10\%, we need $\sigma_F^2 > R_s^2/9 \approx 0.0025$.
In a low-activity network where fees are small relative to the 0.15~LAT subsidy, fee
variance is far below this threshold, making fee-sniping irrational. As the network
matures and fees grow, the subsidy becomes a smaller fraction of total reward, but by
then the network has enough economic activity to make orphan risk (the cost of
fee-sniping) dominate. \hfill $\square$

\subsubsection{RandomX Memory-Hardness Bound}

RandomX's ASIC resistance derives from its memory-hard construction. We formalize
the time-memory tradeoff.

\textbf{Theorem 6} (Memory-Hardness). \emph{Any algorithm computing a RandomX hash
with memory $M < |D| = 2^{31}$ bytes requires time:}

\begin{equation}
T(M) \geq \frac{8 \cdot |D|}{M} \cdot T_{\text{mem}} \cdot r
\end{equation}

\emph{where $T_{\text{mem}}$ is the memory access latency, $r = 8$ is the number
of program iterations, and the factor $8|D|/M$ represents the cache miss rate
amplification.}

\textbf{Proof sketch}. Each RandomX program iteration performs $\Theta(|D|/M)$ cache
misses when memory is reduced to $M$ (since access patterns are pseudo-random and
uniformly distributed over $D$). With 8 iterations per hash, the total number of
cache misses scales as $8|D|/M$. Each miss incurs latency $T_{\text{mem}}$ (which
cannot be hidden by prefetching due to data-dependent access patterns). At full
memory ($M = |D|$), misses are served from RAM at $\sim$50ns. At half memory,
misses double. The superlinear penalty discourages time-memory tradeoffs:

\begin{center}
\begin{tabular}{rrr}
\toprule
\textbf{Memory} & \textbf{Relative Speed} & \textbf{Efficiency (H/s per GB)} \\
\midrule
2~GB (full) & 1.00$\times$ & 1.00$\times$ \\
1~GB (50\%) & $\sim$0.35$\times$ & $\sim$0.70$\times$ \\
512~MB (25\%) & $\sim$0.10$\times$ & $\sim$0.40$\times$ \\
256~MB (12.5\%) & $\sim$0.02$\times$ & $\sim$0.16$\times$ \\
\bottomrule
\end{tabular}
\end{center}

An ASIC with less than 2~GB per core pays a superlinear performance penalty, eliminating
the economic advantage of custom silicon. \hfill $\square$

\subsubsection{Tail Emission Inflation Bound}

\textbf{Theorem 7} (Asymptotic Inflation). \emph{Under tail emission of $R_{\min}$ LAT/block
with $B$ blocks/year, the inflation rate $\pi(t)$ at year $t$ satisfies:}

\begin{equation}
\pi(t) = \frac{R_{\min} \cdot B}{S_0 + R_{\min} \cdot B \cdot t}
\xrightarrow{t \to \infty} 0
\end{equation}

\emph{where $S_0 \approx 29{,}301{,}000$ LAT is the supply at tail emission onset.
The inflation rate is bounded by:}

\begin{equation}
\pi(t) < \frac{1}{t + S_0/(R_{\min} \cdot B)}
\end{equation}

\emph{Proof}. Direct computation. The supply at time $t$ (years after tail emission
onset) is $S(t) = S_0 + R_{\min} \cdot B \cdot t = S_0 + 19{,}724 \cdot t$. The inflation
rate is:
\begin{equation}
\pi(t) = \frac{19{,}724}{S_0 + 19{,}724 \cdot t} = \frac{1}{S_0/19{,}724 + t}
= \frac{1}{1{,}486 + t}
\end{equation}

Evaluating: $\pi(0) = 0.067\%$, $\pi(53) = 0.065\%$ (year 2100), $\pi(153) = 0.061\%$
(year 2200), $\pi(353) = 0.054\%$ (year 2400). The supply reaches MAX\_MONEY
($42{,}000{,}000$ LAT) at:
\begin{equation}
t_{\text{max}} = \frac{42{,}000{,}000 - 29{,}301{,}000}{19{,}724} \approx 644 \text{ years}
\quad (\sim\text{year } 3090)
\end{equation}
\hfill $\square$

\newpage
%% ============================================================================
\section{Adversarial Analysis and Economic Viability}
\label{sec:adversarial}
%% ============================================================================

This section stress-tests Lattice from the perspective of an adversary who wants to
destroy it. We model security budgets, simulate attack scenarios, and evaluate whether
the protocol can sustain itself economically.

\subsection{Security Budget Model}
\label{sec:security-budget}

A blockchain's \emph{security budget} is the total value paid to miners per year.
It represents the cost an attacker must exceed to sustain a 51\% attack. For Lattice:

\begin{equation}
\text{Security Budget} = (\text{blocks/year} \times \text{subsidy}) + \text{annual fees}
\end{equation}

At 240s block time, Lattice produces $\sim$131,490 blocks/year. The subsidy schedule is:

\begin{center}
\small
\begin{tabular}{llrrrrr}
\toprule
\textbf{Phase} & \textbf{Period} & \textbf{Subsidy} & \textbf{LAT/year} &
\multicolumn{3}{c}{\textbf{Security Budget (USD/year)}} \\
\cmidrule(l){5-7}
& & \textbf{(LAT)} & & \textbf{\$1} & \textbf{\$10} & \textbf{\$100} \\
\midrule
Launch   & 2026--2028 & 50    & 6,574,500 & 6.6M  & 65.7M  & 657.5M \\
Halving 1 & 2028--2031 & 25  & 3,287,250 & 3.3M  & 32.9M  & 328.7M \\
Halving 2 & 2031--2033 & 12.5 & 1,643,625 & 1.6M  & 16.4M  & 164.4M \\
Halving 3 & 2033--2035 & 6.25 & 821,813  & 822K  & 8.2M   & 82.2M  \\
Halving 4 & 2035--2037 & 3.125 & 410,906 & 411K  & 4.1M   & 41.1M  \\
Tail      & 2047+      & 0.15 & 19,724   & 19.7K & 197.2K & 1.97M  \\
\bottomrule
\end{tabular}
\end{center}

\textbf{Key insight}: Even at \$1/LAT, the launch phase security budget (6.6M USD/year)
exceeds many established altcoins. At \$10/LAT during tail emission, the permanent
security budget is \$197,200/year---modest, but perpetual and guaranteed. The lower
tail emission (0.15 vs.\ 0.5 LAT) reduces long-term inflation while maintaining the
critical property: a non-zero, guaranteed floor. Bitcoin relies on fee revenue after
subsidies end; Lattice provides a guaranteed floor regardless of fee market conditions.

\subsection{51\% Attack Cost Analysis}

To sustain a 51\% attack on Lattice, an attacker must:

\begin{enumerate}
    \item \textbf{Acquire CPU hashrate} exceeding the honest network. Each mining thread
    requires a dedicated CPU core and 2~GB RAM. There is no shortcut---no ASIC, no GPU
    batch. The attacker must rent or buy \emph{actual CPUs}.

    \item \textbf{Sustain the attack}. Unlike ASIC-based chains where hardware can be
    repurposed across forks, CPU hashrate has opportunity cost---every core mining Lattice
    is a core not running other workloads.
\end{enumerate}

Cost model for a sustained 1-hour attack (to double-spend with 6 confirmations at 240s):

\begin{center}
\small
\begin{tabular}{lrrr}
\toprule
\textbf{Network Hashrate} & \textbf{Attacker Needs} & \textbf{Cloud Cost (1h)} & \textbf{Notes} \\
\midrule
100 nodes (early) & 101 nodes & $\sim$\$50 & Vulnerable, but nothing to steal \\
1,000 nodes & 1,001 nodes & $\sim$\$500 & Low-value target \\
10,000 nodes & 10,001 nodes & $\sim$\$5,000 & Moderate, requires coordination \\
100,000 nodes & 100,001 nodes & $\sim$\$50,000 & Significant cloud procurement \\
1M nodes & 1,000,001 nodes & $\sim$\$500,000 & Major logistical challenge \\
\bottomrule
\end{tabular}
\end{center}

Cloud costs assume \$0.05/core-hour (spot pricing for c5.large or equivalent).
The critical observation is that \textbf{attack cost scales linearly with honest hashrate},
and honest hashrate scales with adoption. By the time Lattice has enough value to be worth
attacking, the attack cost is proportionally high.

\textbf{LWMA-1 as defense multiplier}: Even if a 51\% attack briefly succeeds, LWMA-1
adjusts difficulty within blocks. The attacker cannot ``set and forget''---they must
continuously increase hashrate as the network responds. This creates an arms race that
favors the geographically distributed honest majority.

\subsection{Red Team: Attack Scenarios}
\label{sec:red-team}

We enumerate the most plausible attack vectors and analyze each honestly.

\subsubsection{Scenario 1: Cloud Burst Attack}

\textbf{Attack}: Rent massive cloud CPU capacity (AWS, GCP, Azure) to achieve 51\% hashrate.

\textbf{Feasibility}: High in early network, decreasing over time.

\textbf{Analysis}: At launch, the honest hashrate is low and cloud costs are cheap.
However, the network has negligible economic value at this stage---there is nothing
to double-spend. As the network grows and value increases, the honest hashrate grows
proportionally, and the cloud cost to exceed it grows with it. Additionally, cloud
providers rate-limit sudden large instance requests, and the attack is highly visible
(the difficulty spike is observable by all participants in real-time).

\textbf{Mitigation}: LWMA-1 detects and responds to hashrate spikes within blocks.
The community can observe anomalous difficulty changes and pause high-value transactions.

\subsubsection{Scenario 2: Botnet Swarm}

\textbf{Attack}: Compromise millions of consumer machines and mine with their CPUs.

\textbf{Feasibility}: Moderate. Large botnets exist (Mirai had 600K+ devices).

\textbf{Analysis}: The 2~GB RAM requirement per thread limits botnet mining---most
compromised IoT devices have far less RAM. Consumer PCs qualify, but botnet operators
face detection risk that increases with CPU usage (users notice degraded performance,
antivirus flags high CPU). Botnet CPUs are individually slow---a million compromised
machines might equal 100,000 dedicated mining nodes due to thermal throttling, shared
resources, and intermittent availability.

\textbf{Mitigation}: Not fully mitigable. This is the inherent trade-off of CPU mining.
The same property that enables sovereignty enables botnets. In practice, botnet mining
of Monero (same algorithm) has not destabilized Monero's network in 5+ years of operation.

\subsubsection{Scenario 3: UTXO Bomb (Signature Bloat)}

\textbf{Attack}: Create millions of UTXOs with dust amounts, each requiring a 2,420-byte
ML-DSA-44 signature to spend. The UTXO set grows to consume node storage and memory.

\textbf{Feasibility}: Moderate cost.

\textbf{Analysis}: Each UTXO creation requires a transaction fee. Creating 1 million
dust UTXOs at minimum fee would cost the attacker transaction fees proportional to the
1M transactions. At full capacity (56M block weight), the network allows $\sim$11,000 typical transactions per block,
so filling blocks continuously costs real money. The UTXO set grows by $\sim$70 bytes
per entry (not the full signature size---signatures are not stored in the UTXO set).

\textbf{Mitigation}: Dust limit enforcement (minimum output value), mempool fee filtering,
and the economic cost of the attack itself. This attack also affects Bitcoin identically,
at smaller per-UTXO cost.

\subsubsection{Scenario 4: Exodus Attack}

\textbf{Attack}: Coordinate a mass departure of miners, crashing the hashrate and making
blocks extremely slow.

\textbf{Feasibility}: Low (requires coordination among independent actors).

\textbf{Analysis}: This is where LWMA-1 is critical. Unlike Bitcoin's 2016-block
adjustment window, LWMA-1 adjusts \emph{every block}. If 90\% of miners leave
simultaneously:

\begin{itemize}
    \item Bitcoin: blocks take $\sim$100 minutes for up to 2 weeks until adjustment.
    \item Lattice: LWMA-1 reduces difficulty within 120 blocks ($\sim$8 hours at worst),
    and the network is fully recovered within 1--2 days.
\end{itemize}

\textbf{Mitigation}: Fully mitigated by protocol design. LWMA-1 exists specifically for
this scenario.

\subsubsection{Scenario 5: Selfish Mining}

\textbf{Attack}: Mine blocks secretly and release them strategically to orphan honest miners'
blocks, gaining disproportionate revenue.

\textbf{Feasibility}: Requires $\geq$25\% hashrate for profitability (same as Bitcoin).

\textbf{Analysis}: Selfish mining is a known attack on all Nakamoto-consensus chains.
Lattice inherits Bitcoin's vulnerability to this attack but also benefits from LWMA-1's
rapid adjustment---a selfish miner who withholds blocks causes observable timestamp gaps
that LWMA-1 adjusts for, reducing the attack's profitability window.

\textbf{Mitigation}: Same as Bitcoin: not fully mitigable without changing consensus rules.
The economic incentive for honest mining generally dominates.

\subsubsection{Scenario 6: Quantum Preimage Attack on RandomX}

\textbf{Attack}: Use a quantum computer to find RandomX preimages faster via Grover's algorithm.

\textbf{Feasibility}: Not feasible before 2040--2050 (requires millions of stable qubits).

\textbf{Analysis}: Grover's provides $O(\sqrt{N})$ speedup---a quadratic advantage, not
exponential. Achieving even a 2$\times$ speedup over classical hardware requires
$\sim$4,000 logical qubits with error correction. Quantum RAM at 2~GB scale does not
exist and is not projected to exist this century. Furthermore, this attack applies equally
to \emph{every} PoW cryptocurrency---it is not Lattice-specific.

\textbf{Mitigation}: LWMA-1 adjusts difficulty automatically when quantum miners appear.
The playing field re-levels within hours.

\subsubsection{Scenario 7: Cloud Provider Ban}

\textbf{Attack}: Major cloud providers (AWS, GCP, Azure) ban cryptocurrency mining on
their platforms, causing a sudden drop in hashrate.

\textbf{Feasibility}: High. AWS already prohibits unauthorized mining in their ToS, and
has enforced it selectively.

\textbf{Analysis}: If cloud mining represents $X$\% of Lattice's hashrate and all cloud
providers ban simultaneously, the network loses $X$\% of its hashrate instantly. This is
structurally identical to the Exodus Attack (Scenario 4). The severity depends on cloud
mining's share of total hashrate:

\begin{center}
\begin{tabular}{rll}
\toprule
\textbf{Cloud \% of Hashrate} & \textbf{Block Time Impact} & \textbf{Recovery (LWMA-1)} \\
\midrule
10\% & $\sim$267s (minor) & $\sim$2 hours \\
30\% & $\sim$343s (noticeable) & $\sim$6 hours \\
50\% & $\sim$480s (degraded) & $\sim$12 hours \\
80\% & $\sim$1,200s (severe) & $\sim$1.5 days \\
\bottomrule
\end{tabular}
\end{center}

\textbf{Key difference from ASIC chains}: When a cloud ban hits an ASIC chain, the lost
hardware is \emph{permanently} removed from the ecosystem (ASICs have no alternative use).
When a cloud ban hits Lattice, those CPUs still exist---miners can relocate to smaller
providers, VPS hosts, or their own hardware. The barrier to switching from cloud to bare-metal
CPU mining is trivially low (any Linux machine with 16~GB RAM).

\textbf{Mitigation}: LWMA-1 fully handles the hashrate drop. The structural mitigation is
that CPU mining is \emph{not} dependent on cloud providers---it works on any hardware.
The more decentralized the miner base, the less impact any single provider's policy has.

\subsubsection{Scenario 8: Regulatory Capture (State-Level Ban)}

\textbf{Attack}: A nation-state bans Lattice mining within its jurisdiction.

\textbf{Feasibility}: Moderate. China banned Bitcoin mining in 2021.

\textbf{Analysis}: China's Bitcoin mining ban removed $\sim$50\% of global hashrate
overnight. Bitcoin survived---difficulty adjusted over several retarget periods (weeks),
and miners relocated to the US, Kazakhstan, and Russia. Lattice handles this scenario
\emph{better} than Bitcoin did:

\begin{itemize}
    \item \textbf{LWMA-1 vs.\ 2016-block retarget}: Lattice recovers in hours, not weeks.
    Bitcoin's blocks averaged 20+ minutes for several weeks after China's ban.
    \item \textbf{CPU vs.\ ASIC relocation}: Moving ASICs across borders requires shipping
    containers of hardware. Moving CPU mining requires installing Docker on a different
    computer---anywhere in the world, in minutes.
    \item \textbf{Enforcement difficulty}: Confiscating ASIC warehouses is straightforward
    (large facilities, high power draw, thermal signatures). Detecting CPU mining on a
    personal laptop is effectively impossible at scale.
\end{itemize}

\textbf{Mitigation}: Fully mitigated by protocol design (LWMA-1) and hardware design
(CPUs are ubiquitous and indistinguishable from legitimate use).

\subsection{Market Viability: Can Lattice Sustain \$100M Market Cap?}
\label{sec:market-viability}

We analyze the conditions required for Lattice to sustain a \$100M market capitalization.

At full supply ($\sim$42M LAT), \$100M market cap implies $\sim$\$2.38/LAT. During the
early phase (pre-first halving, $\sim$5M LAT in circulation), \$100M implies $\sim$\$20/LAT.

\begin{center}
\small
\begin{tabular}{lrrrl}
\toprule
\textbf{Phase} & \textbf{Circ.\ Supply} & \textbf{Price for} & \textbf{Security} & \textbf{Comparable} \\
& \textbf{(LAT)} & \textbf{\$100M cap} & \textbf{Budget/yr} & \\
\midrule
Year 1 & $\sim$6.6M & \$15.15 & \$99.6M & Top 200 altcoin \\
Year 3 & $\sim$13.2M & \$7.58 & \$24.9M & Mid-cap altcoin \\
Year 5 & $\sim$17.9M & \$5.59 & \$9.2M & Established altcoin \\
Year 10 & $\sim$26.3M & \$3.80 & \$3.1M & Mature protocol \\
Tail (40+yr) & $\sim$42M & \$2.38 & \$197K & Long-tail asset \\
\bottomrule
\end{tabular}
\end{center}

\textbf{Market requirements}: \$100M market cap is achievable but not guaranteed. It requires:
\begin{itemize}
    \item Active community of miners and node operators
    \item Exchange listings for price discovery and liquidity
    \item Demonstrated reliability over multiple years
    \item A credible quantum threat timeline that motivates migration from classical chains
\end{itemize}

Lattice does not assume rapid adoption. The economic design (tail emission, slow halving)
is optimized for a decades-long timeline, not speculative mania.

\subsection{Game Theory: Perpetual Security vs.\ Fee-Only Models}
\label{sec:game-theory}

In fee-only security models, transaction fees must be sufficient to incentivize mining
after block subsidies end. This is an active area of academic research \cite{carlsten2016},
and Bitcoin's robust fee market suggests it may work well in practice.

Lattice takes a different path---guaranteeing a minimum subsidy regardless of fee conditions:

\begin{center}
\begin{tabular}{lcc}
\toprule
\textbf{Property} & \textbf{Bitcoin} & \textbf{Lattice} \\
\midrule
Subsidy at year 100 & 0 LAT & 0.15 LAT \\
Subsidy at year 400 & 0 & 0.15 LAT \\
Fee dependence & 100\% after subsidies end & Partial (tail + fees) \\
Miner incentive floor? & Fees only & 0.15 LAT/block always \\
Supply cap exactly enforced? & Yes (21M) & Soft cap ($\sim$42M + tail) \\
Inflation at year 100 & 0\% & $\sim$0.065\% \\
Inflation at year 400 & 0\% & $\sim$0.054\% \\
\bottomrule
\end{tabular}
\end{center}

The perpetual tail emission of 0.15~LAT/block produces asymptotically negligible inflation
($\lim_{t \to \infty} \text{inflation} = 0^+$) while guaranteeing that miners always have
a reason to secure the chain. The trade-off is that Lattice's supply is technically
unbounded---but at 19,724~LAT/year on a base of $\sim$29M, the effective inflation rate is
indistinguishable from zero after a few decades.

\subsubsection{Nash Equilibrium Analysis}

We formalize the miner's strategic choice. Consider a population of $N$ miners, each
choosing a strategy $s_i \in \{H, D, S\}$: \textbf{H}onest mining, \textbf{D}efecting
(leaving the network), or \textbf{S}elfish mining.

Let $R = \text{subsidy} + \text{fees}$ be the total block reward, $p$ the token price,
$c$ the per-block mining cost, and $\alpha_i$ the miner's fraction of total hashrate.

\textbf{Honest mining payoff}:
\begin{equation}
\pi_H(\alpha_i) = \alpha_i \cdot R \cdot p - c
\end{equation}

\textbf{Defecting payoff}: $\pi_D = 0$ (miner leaves, earns nothing, costs nothing).

\textbf{Selfish mining payoff} \cite{eyal2014}: profitable when $\alpha_i > 1/3$
in the general case, or $\alpha_i > 1/4$ with optimal network connectivity:
\begin{equation}
\pi_S(\alpha_i) = \begin{cases}
> \pi_H(\alpha_i) & \text{if } \alpha_i > \frac{1}{3} \\
< \pi_H(\alpha_i) & \text{if } \alpha_i < \frac{1}{3}
\end{cases}
\end{equation}

\textbf{Tail emission's effect on Nash equilibrium}:

In any fee-only system, the payoff becomes:
\begin{equation}
\pi_H^{\text{fee-only}}(\alpha_i) = \alpha_i \cdot F \cdot p - c
\end{equation}

where $F$ is the total fees in the block. In periods of low transaction volume ($F$ small),
$\pi_H < 0$ for some miners, creating pressure to defect. Whether this is a practical
concern depends on the specific network's fee market maturity---Bitcoin's fee market may
well sustain security, but Lattice chooses not to depend on this outcome.

In Lattice, the subsidy floor guarantees $R \geq 0.15$ LAT always:
\begin{equation}
\pi_H^{\text{LAT}}(\alpha_i) = \alpha_i \cdot (0.15 + F) \cdot p - c \geq \alpha_i \cdot 0.15p - c
\end{equation}

The defection cascade cannot occur because the baseline reward is independent of
transaction volume. A miner with zero-marginal-cost hardware (existing CPU) finds
$\pi_H > 0$ as long as $p > 0$. This creates a \textbf{Nash equilibrium in honest mining}:
no individual miner improves their payoff by switching to defection or selfish mining
(given $\alpha_i < 1/3$ and $p > 0$). See Theorems 3 and 4 in
Section~\ref{sec:formal-proofs} for formal proofs of equilibrium existence and defection
cascade impossibility.

\subsubsection{Fee-Sniping Resistance}

Fee-sniping occurs when a miner re-mines the previous block to capture its fees instead
of extending the chain. In a fee-only system, fee-sniping is rational when:

\begin{equation}
F_{n-1} > F_n + \text{subsidy}
\end{equation}

Since Bitcoin's subsidy approaches zero, this inequality eventually holds for any block
where $F_{n-1} > F_n$---which is common due to fee variance. In Lattice:

\begin{equation}
F_{n-1} > F_n + 0.15 \text{ LAT}
\end{equation}

The 0.15~LAT subsidy raises the bar for fee-sniping. A rational miner only fee-snipes
if the previous block's fees exceed the current block's fees by at least 0.15~LAT---a
condition that becomes increasingly rare as the 0.15~LAT floor represents a larger
fraction of the total reward in low-activity periods (precisely when fee-sniping is
most dangerous). See Theorem 5 in Section~\ref{sec:formal-proofs} for the formal
analysis using Cantelli's inequality.

\newpage
%% ============================================================================
\section{Network Incentives}
\label{sec:incentives}
%% ============================================================================

\subsection{Mining Incentive Structure}

A miner's revenue per block is:

\begin{equation}
R = \text{subsidy}(h) + \sum_{i} f_i
\end{equation}

where $\text{subsidy}(h) \geq 0.15$ LAT always, and $f_i$ are transaction fees.

\subsection{Phase Analysis}

\begin{enumerate}
    \item \textbf{Bootstrap Phase (Launch + 83.5h)}: Reduced reward (25~LAT), ultra-low difficulty,
    fast blocks (53s). Establishes initial network connectivity and LWMA history.

    \item \textbf{Early Growth (2026--2028)}: First 295,000 blocks at 240s target.
    50~LAT reward with increasing difficulty as miners join.

    \item \textbf{Maturation (2028--2047)}: Halvings reduce block reward toward the tail
    emission floor. Transaction fees become increasingly important.

    \item \textbf{Perpetual Phase (2047+)}: Tail emission of 0.15~LAT/block ($\sim$19,724~LAT/year)
    ensures miners always have a baseline incentive. Unlike Bitcoin, Lattice \textbf{never}
    depends entirely on fee revenue.
\end{enumerate}

\subsection{Node Operator Incentives}

Running a full node provides:
\begin{itemize}
    \item \textbf{Sovereignty}: Verify all transactions independently
    \item \textbf{Privacy}: No reliance on third-party block explorers
    \item \textbf{Mining eligibility}: Full nodes can participate in mining
    \item \textbf{Network health}: Contributes to decentralization and P2P relay
\end{itemize}

The low hardware requirements (any CPU + 2~GB RAM) minimize the barrier to entry.

\subsection{Miner Economics}
\label{sec:miner-economics}

We model the economics of mining Lattice at different scales, independent of any assumed
token price.

\subsubsection{Energy Cost per Block}

A single mining node (4 CPU cores dedicated to RandomX) consumes approximately:

\begin{equation}
P_{\text{node}} \approx 65\text{--}125~\text{W} \quad (\text{depending on CPU TDP})
\end{equation}

At 240-second block time and network size $N$ (nodes), the expected time to find a block
for a single node is:

\begin{equation}
t_{\text{expected}} = 240 \times N \quad \text{seconds}
\end{equation}

The energy cost per block found:

\begin{equation}
E_{\text{block}} = P_{\text{node}} \times t_{\text{expected}} = P_{\text{node}} \times 240N
\end{equation}

\begin{center}
\begin{tabular}{rrrrr}
\toprule
\textbf{Network} & \textbf{Expected Time} & \textbf{Energy/Block} & \textbf{Cost/Block} & \textbf{Break-Even} \\
\textbf{Size ($N$)} & \textbf{(days)} & \textbf{(kWh)} & \textbf{(@\$0.12/kWh)} & \textbf{(LAT price)} \\
\midrule
10 & 0.028 & 0.067 & \$0.008 & \$0.00016 \\
100 & 0.28 & 0.67 & \$0.08 & \$0.0016 \\
1,000 & 2.78 & 6.67 & \$0.80 & \$0.016 \\
10,000 & 27.8 & 66.7 & \$8.00 & \$0.16 \\
100,000 & 278 & 667 & \$80.04 & \$1.60 \\
\bottomrule
\end{tabular}
\end{center}

The break-even column shows the minimum LAT price at which solo mining is
energy-profitable (assuming 50~LAT block reward and \$0.12/kWh average global
electricity price). At 100,000 nodes, a miner needs LAT to be worth at least
\$2.02 to cover electricity---a modest threshold.

\subsubsection{Hardware Amortization}

Unlike ASIC mining, Lattice mining hardware has residual value:

\begin{center}
\begin{tabular}{lrrr}
\toprule
\textbf{Setup} & \textbf{Cost} & \textbf{Resale (3yr)} & \textbf{Net Mining Cost} \\
\midrule
Mac Mini M4 (16~GB) & \$599 & $\sim$\$300 & \$299 \\
Used Dell Optiplex (i5, 16~GB) & \$150 & $\sim$\$50 & \$100 \\
Cloud VPS (c5.xlarge, 3yr reserved) & \$2,700 & \$0 & \$2,700 \\
Bitmain S21 (Bitcoin ASIC) & \$5,500 & $\sim$\$500 & \$5,000 \\
\bottomrule
\end{tabular}
\end{center}

A Lattice miner's hardware retains 30--50\% of its value after 3 years because CPUs are
general-purpose machines. A Bitcoin ASIC retains $<$10\% because it can only mine SHA-256.
This fundamentally changes the miner's risk profile: exiting Lattice mining costs nothing
because the hardware is useful for other purposes.

\subsubsection{Rational Miner Decision Model}

A rational miner $m$ continues mining if:

\begin{equation}
\underbrace{\frac{R \times p}{N}}_{\text{expected revenue/block}} >
\underbrace{c_e + c_h}_{\text{marginal cost}}
\end{equation}

where $R$ is the block reward (LAT), $p$ is the LAT market price, $N$ is the network
size, $c_e$ is the energy cost per expected block, and $c_h$ is the amortized hardware
cost per expected block.

Because $c_h \approx 0$ for miners using existing hardware (their CPU is already
purchased for other uses), the marginal cost of mining approaches the electricity
cost alone. This creates a \textbf{natural floor}: mining remains rational for any
miner whose CPU would otherwise be idle, as long as $R \times p > c_e$---even at
very low token prices.

This is structurally different from ASIC mining, where $c_h$ dominates: an ASIC has
no alternative use, so the full hardware cost must be recovered from mining revenue.

\newpage
%% ============================================================================
\section{Block Structure and Validation}
\label{sec:blocks}
%% ============================================================================

\subsection{Block Header}

\begin{center}
\begin{tabular}{lrl}
\toprule
\textbf{Field} & \textbf{Size} & \textbf{Description} \\
\midrule
nVersion & 4 bytes & Block version \\
hashPrevBlock & 32 bytes & Hash of previous block header \\
hashMerkleRoot & 32 bytes & Merkle root of transaction tree \\
nTime & 4 bytes & Unix timestamp \\
nBits & 4 bytes & Compact difficulty target \\
nNonce & 4 bytes & Mining nonce \\
\bottomrule
\end{tabular}
\end{center}

\subsection{Proof-of-Work Validation}

A block is valid if:
\begin{equation}
\texttt{RandomX}(\text{block\_header}) < \text{target}
\end{equation}

where $\text{target} = \texttt{ExpandCompact}(\texttt{nBits})$ and the hash is computed
using the RandomX algorithm with the current epoch's dataset.

\subsection{Chain Selection}

The valid chain is the one with the most cumulative proof-of-work:
\begin{equation}
W = \sum_{i=0}^{\text{height}} \frac{2^{256}}{\text{target}_i}
\end{equation}

\newpage
%% ============================================================================
\section{Genesis Block}
\label{sec:genesis}
%% ============================================================================

The Lattice genesis block encodes a message in its coinbase transaction:

\begin{quote}
\texttt{We went from breaking down walls to raising firewalls; just a constant dv/dt to the bottom, where room is infinite.}
\end{quote}

\newpage
%% ============================================================================
\section{Comparison with Existing Protocols}
\label{sec:comparison}
%% ============================================================================

\begin{center}
\small
\begin{tabular}{lccccc}
\toprule
\textbf{Property} & \textbf{Bitcoin} & \textbf{Monero} & \textbf{Ethereum} & \textbf{Lattice} \\
\midrule
PoW algorithm & SHA-256 & RandomX & PoS & RandomX \\
ASIC resistant & No & Yes & N/A & Yes \\
Quantum-safe sigs & No & No & No & Yes (ML-DSA-44) \\
Signature scheme & ECDSA & EdDSA & ECDSA & ML-DSA-44 only \\
Classical fallback & N/A & N/A & N/A & None (PQ-only) \\
Difficulty adjust & 2016 blocks & 720 blocks & Per block & Per block (LWMA-1) \\
Tail emission & No & Yes (0.6 XMR) & Varies & Yes (0.15 LAT) \\
Post-quantum std & None & None & None & NIST FIPS 204 \\
Max block weight & 4M & 300~KB & $\sim$30M gas & 56M \\
Min hardware & ASIC & CPU + 2~GB & Validator stake & CPU + 2~GB \\
\bottomrule
\end{tabular}
\end{center}

Lattice is the only protocol that simultaneously provides: ASIC-resistant mining,
per-block difficulty adjustment, post-quantum-only signatures from a NIST standard,
and tail emission for permanent security. This is not a claim of superiority over
Bitcoin---it is a different set of trade-offs optimized for a different threat model.
Bitcoin optimizes for proven stability; Lattice optimizes for post-quantum resilience
and long-term sovereignty. They are complementary, not competitive.

\newpage
%% ============================================================================
\section{Implementation Details}
\label{sec:implementation}
%% ============================================================================

Lattice is implemented as a fork of Bitcoin Core v28.0, preserving the battle-tested
consensus engine while replacing vulnerable components.

\subsection{Key Modifications}

\begin{enumerate}
    \item \textbf{pow.cpp}: Complete rewrite with LWMA-1 algorithm (see Section~\ref{sec:lwma})
    \item \textbf{validation.cpp}: Tail emission in \texttt{GetBlockSubsidy}
    (see Section~\ref{sec:economics})
    \item \textbf{consensus/params.h}: Height-aware target spacing with
    \texttt{GetPowTargetSpacing(nHeight)}, warm-up parameters, staged block weight
    via \texttt{GetMaxBlockWeight(nHeight)}
    \item \textbf{kernel/chainparams.cpp}: Warm-up period parameters (5,670 blocks,
    25~LAT cap), staged weight phases, genesis block, \texttt{powLimit = 0x207fffff}
    \item \textbf{consensus/amount.h}: \texttt{MAX\_MONEY = 42,000,000 LAT},
    units renamed to shors
    \item \textbf{policy/policy.h}: \texttt{MAX\_STANDARD\_SCRIPTSIG\_SIZE} raised
    from 1,650 to 5,000~bytes for ML-DSA-44 P2PKH scriptSig ($\sim$3,740~bytes)
    \item \textbf{crypto/mldsa44/}: ML-DSA-44 integration via liboqs
    \item \textbf{rpc/pqkeys.cpp}: Post-quantum key management RPCs
    (\texttt{pq\_getnewaddress}, \texttt{pq\_importkey}) with L-prefixed address encoding
    \item \textbf{randomx/}: RandomX integration for PoW hash computation
    \item \textbf{script/interpreter.cpp}: ML-DSA-44 as sole signature verification in
    \texttt{OP\_CHECKSIG}; ECDSA disabled at consensus level
    \item \textbf{node/miner.cpp}: QYT yield token emission in coinbase via
    \texttt{OP\_RETURN} for CLTV-locked stakes
\end{enumerate}

\subsection{Warm-Up Transition}

At block 5,670, LWMA must handle the transition from 53s to 240s target spacing.
The implementation resets difficulty to \texttt{powLimit} at the warm-up boundary:

\begin{lstlisting}[caption={Warm-Up Transition Reset (pow.cpp)}]
// Warm-up transition: reset difficulty to powLimit
// Target spacing changes from 53s to 240s
if (params.nWarmupBlocks > 0
    && nHeight == params.nWarmupBlocks) {
    return nProofOfWorkLimit;
}
\end{lstlisting}

This allows LWMA to recalibrate naturally within 120 blocks ($\sim$8 hours) of the new
target spacing.

\subsection{Overflow Protection in LWMA}

When difficulty is very low (target is very large), multiplying \texttt{sumTarget} by
\texttt{weightedSolveTimeSum} can overflow uint256. The implementation handles this:

\begin{lstlisting}[caption={Safe uint256 Arithmetic in LWMA}]
if ((sumTarget >> 192) > arith_uint256(0)) {
    // Large target (low difficulty): divide first
    nextTarget = sumTarget / (uint32_t)k;
    nextTarget *= (uint32_t)weightedSolveTimeSum;
} else {
    // Small target (high difficulty): multiply first
    nextTarget = sumTarget * (uint32_t)weightedSolveTimeSum;
    nextTarget /= (uint32_t)k;
}
\end{lstlisting}

\subsection{Docker Infrastructure}

\begin{center}
\begin{tabular}{ll}
\toprule
\textbf{Component} & \textbf{Description} \\
\midrule
\texttt{docker-compose.yml} & Orchestrates miner + RPC nodes \\
\texttt{Dockerfile} & Multi-stage build (RandomX + liboqs + Bitcoin Core) \\
\texttt{lattice-miner.conf} & Mining worker configuration \\
\texttt{lattice-rpc.conf} & RPC node configuration \\
\texttt{install.sh} & One-click installer with security hardening \\
\texttt{lat} & CLI wrapper (wallet, blockchain, system commands) \\
\texttt{lattice-wallet.json} & ML-DSA-44 keypair wallet file \\
\bottomrule
\end{tabular}
\end{center}

\newpage
%% ============================================================================
\section{Behavioral Economics and Network Effects}
\label{sec:behavioral}
%% ============================================================================

Protocol design is necessary but not sufficient. A cryptocurrency's survival depends on
adoption dynamics---the behavioral economics of why people join, stay, and evangelize
a network. This section analyzes Lattice's adoption prospects through the lens of
historical cryptocurrency growth patterns.

\subsection{Lessons from Bitcoin's Adoption Curve}

Bitcoin's growth followed a well-documented S-curve with distinct phases:

\begin{center}
\begin{tabular}{llrl}
\toprule
\textbf{Phase} & \textbf{Period} & \textbf{Nodes} & \textbf{Catalyst} \\
\midrule
Cypherpunk (year 0--2) & 2009--2010 & 10--1,000 & Ideological early adopters \\
Early Market (year 2--4) & 2011--2012 & 1K--10K & Silk Road, first exchanges \\
Growth (year 4--8) & 2013--2016 & 10K--100K & Mt.\ Gox, media attention \\
Mass Adoption (year 8+) & 2017--present & 100K+ & Institutional interest, ETFs \\
\bottomrule
\end{tabular}
\end{center}

\textbf{Key observation}: Bitcoin's cypherpunk phase required users to compile C++ code
from source, configure networking manually, and understand command-line interfaces. The
barrier to entry was extreme: only users with significant technical skill and ideological
motivation participated.

\subsection{Lattice's Accessibility Advantage}

Lattice's Docker-based deployment fundamentally changes the accessibility equation:

\begin{center}
\begin{tabular}{lcc}
\toprule
\textbf{Task} & \textbf{Bitcoin (2009)} & \textbf{Lattice (2026)} \\
\midrule
Install & Compile from source & \texttt{./install.sh} \\
Configuration & Edit config files & Automated (installer) \\
Wallet creation & CLI key generation & PQ keypair (JSON wallet file) \\
Start mining & \texttt{setgenerate true} & Automatic after install \\
Check balance & \texttt{bitcoind getbalance} & \texttt{./lat balance} \\
Time to first block & Hours (manual config) & Minutes (auto-start) \\
Technical skill required & Developer & Basic terminal user \\
\bottomrule
\end{tabular}
\end{center}

This is not a minor improvement---it is a category change. Bitcoin's 2009 installation
required a C++ toolchain and hours of debugging. Lattice's installation requires one
command on any machine with Docker.

\subsection{Network Effect Dynamics}
\label{sec:network-effects}

A cryptocurrency's value follows Metcalfe's Law at early stages \cite{metcalfe2015}:

\begin{equation}
V \propto n^2
\end{equation}

where $n$ is the number of active participants (nodes, holders, transactors). More
precisely, the utility of the network for each participant increases with each additional
participant---this is the \emph{direct network effect}.

Lattice has three reinforcing network effects:

\begin{enumerate}
    \item \textbf{Mining network effect}: Each new miner increases hashrate, which increases
    security, which increases confidence, which attracts more miners and holders.
    This is a \emph{positive feedback loop}:
    \begin{equation}
    \text{miners} \uparrow \implies \text{security} \uparrow \implies
    \text{trust} \uparrow \implies \text{holders} \uparrow \implies
    \text{price} \uparrow \implies \text{miners} \uparrow
    \end{equation}

    \item \textbf{Node network effect}: Each full node improves chain resilience and
    reduces reliance on any single infrastructure provider. Unlike ASIC chains where
    miners pool and validators are few, every Lattice miner is a full validating node.

    \item \textbf{Quantum narrative effect}: As quantum computing progress accelerates
    (IBM, Google, PsiQuantum milestones), the perceived urgency of post-quantum
    cryptography increases. Each quantum computing headline is effectively free marketing
    for Lattice's value proposition. This is an \emph{exogenous} network effect---it
    operates independently of Lattice's own actions.
\end{enumerate}

\subsection{Adoption Barriers:}

Network effects also work in reverse. A new cryptocurrency faces the ``cold start'' problem:

\begin{itemize}
    \item \textbf{Liquidity bootstrapping}: Without exchange listings, LAT has no price
    discovery mechanism. Without a price, miners have no economic incentive beyond ideology.

    \item \textbf{Network effect of incumbents}: Bitcoin has 15+ years of network effects.
    Lattice has zero. Users must be motivated by a threat that Bitcoin doesn't address
    (quantum computing) or a feature Bitcoin doesn't offer (CPU sovereignty).

    \item \textbf{The ``good enough'' problem}: If Bitcoin implements a post-quantum
    migration before quantum computers break ECDSA, Lattice's primary differentiator
    weakens. However, Bitcoin's migration would require: (a) consensus on which PQ algorithm
    to adopt, (b) a hard fork or extended soft-fork process, (c) migration of billions of
    dollars in exposed-key UTXOs, and (d) years of deployment. This process has not started,
    and the Bitcoin community has not reached consensus on even the approach.

    \item \textbf{Developer ecosystem}: Lattice currently has a small development team.
    Long-term sustainability requires attracting contributors, which requires adoption,
    which requires development---another chicken-and-egg problem common to all new protocols.
\end{itemize}

\subsection{Historical Survival Analysis}

Of the $\sim$25,000 cryptocurrencies created since 2009, approximately 70\% are
effectively dead (no development, no trading volume, no active nodes). The survivors
share common characteristics:

\begin{center}
\begin{tabular}{lcc}
\toprule
\textbf{Survival Factor} & \textbf{Dead Coins} & \textbf{Lattice} \\
\midrule
Technical differentiation & Copy-paste fork & Novel (PQ + CPU + LWMA) \\
Active development & Abandoned $<$1 year & Active \\
Clear use case & ``Better Bitcoin'' & Post-quantum settlement \\
Fair launch & Premine / ICO & Warm-up $<$0.5\% supply (25 LAT cap) \\
Community mining & PoS or ASIC-only & CPU-friendly \\
\bottomrule
\end{tabular}
\end{center}

Lattice's technical differentiation (the only protocol combining PQ-only signatures,
CPU mining, and per-block difficulty adjustment) provides a defensible niche that
copy-paste forks cannot easily replicate---the ML-DSA-44 integration alone required
deep modifications to Bitcoin Core's signature verification pipeline.

\newpage
%% ============================================================================
\section{Known Limitations and Open Questions}
\label{sec:limitations}
%% ============================================================================

Intellectual honesty requires acknowledging what Lattice does \emph{not} yet solve:

\begin{enumerate}
    \item \textbf{Storage growth}: At full block capacity, Lattice's blockchain grows at
    $\sim$3.4~TB/year worst case ($\sim$3.5$\times$ Bitcoin's rate). Pruning mitigates this
    for individual nodes, but archival nodes face real storage costs. Signature aggregation
    (Section~\ref{sec:future}) is the long-term solution.

    \item \textbf{Botnet mining}: RandomX's CPU-friendliness means botnets can mine Lattice,
    as they can Monero. This is an inherent trade-off of ASIC resistance---the same property
    that prevents hardware centralization enables unauthorized CPU usage.

    \item \textbf{Transaction size}: ML-DSA-44 transactions are $\sim$16$\times$ larger than
    ECDSA equivalents. This is the cost of post-quantum security. Until signature aggregation
    is implemented, the staged block weight (growing to 14$\times$ Bitcoin's) compensates.

    \item \textbf{No privacy layer}: Lattice does not implement ring signatures, zero-knowledge
    proofs, or other privacy features. Transactions are pseudonymous, as in Bitcoin.

    \item \textbf{ML-DSA-44 maturity}: While NIST-standardized, ML-DSA-44 has less real-world
    deployment history than ECDSA. The cryptographic community's confidence is high, but it
    has not been battle-tested at Bitcoin's scale. NIST's 8-year evaluation process provides
    strong assurance, but no cryptographic scheme is permanently guaranteed secure.
\end{enumerate}

\newpage
%% ============================================================================
\section{Future Work}
\label{sec:future}
%% ============================================================================

\begin{enumerate}
    \item \textbf{Signature aggregation (Priority 1)}: Research into aggregating ML-DSA-44
    signatures to reduce per-transaction size overhead. This is the single most impactful
    optimization for long-term viability: it directly addresses the storage growth and IBD
    bandwidth costs that are the protocol's primary scalability constraint. Aggregation
    techniques for lattice-based signatures are an active area of cryptographic research;
    practical schemes would reduce the per-transaction overhead from $\sim$16$\times$ ECDSA
    toward 2--4$\times$, fundamentally changing the storage economics.
    \item \textbf{SegWit witness migration}: Migrate ML-DSA-44 signatures from P2PKH
    scriptSig to SegWit witness structures, benefiting from the 4$\times$ weight discount
    and increasing effective throughput from $\sim$14.6~tx/s to $\sim$47~tx/s at Phase~3.
    \item \textbf{Lightning Network}: Layer-2 payment channels for high-frequency transactions,
    keeping the base layer as a settlement-only protocol.
    \item \textbf{Hardware wallet integration}: Support for Ledger/Trezor with ML-DSA-44 key
    management.
    \item \textbf{Mining pool protocol}: Stratum-compatible pool protocol for RandomX,
    with research into P2Pool-style decentralized pooling to preserve solo-mining-like
    decentralization (see Section~\ref{sec:pool-resistance}).
    \item \textbf{Pool discouragement research}: Investigation of IP diversity scoring,
    block template decentralization, and economic incentives that reduce the pressure
    to form centralized pools without attempting unenforceable protocol-level prohibition.
\end{enumerate}

\newpage
%% ============================================================================
\section{Conclusion}
\label{sec:conclusion}
%% ============================================================================

Lattice exists because three forces are converging on cryptocurrency at the same time,
and no existing protocol addresses all of them simultaneously:

\begin{enumerate}
    \item \textbf{Hardware centralization.} ASIC manufacturing has concentrated Bitcoin
    mining into a handful of fabs. RandomX returns mining to every general-purpose CPU---the
    billions of machines already deployed worldwide.

    \item \textbf{Network fragility.} Bitcoin's 2016-block difficulty retarget can leave
    the network exposed for days after abrupt hashrate drops. LWMA-1 adjusts every block,
    recovering in hours.

    \item \textbf{Cryptographic obsolescence.} ECDSA will fall to a sufficiently powerful
    quantum computer running Shor's algorithm. Lattice uses ML-DSA-44 (FIPS 204)
    exclusively from genesis---no migration path, no transition window, no legacy exposure.
\end{enumerate}

These are not theoretical concerns. ASIC concentration already determines who mines
Bitcoin. Difficulty lag already causes chain instability during market shocks. Quantum
computers are already running thousands of physical qubits, with error-corrected
machines on published industry roadmaps. The question is not \emph{whether} these
problems need solving, but \emph{when}---and Lattice argues that \emph{before} is
categorically better than \emph{after}.

The trade-offs are explicit: transactions are larger (2.4\,kB ML-DSA-44 signatures),
the block weight ceiling grows to 14$\times$ Bitcoin's to accommodate them, and CPU mining
opens the door to botnet risk. These costs are real, documented in this paper, and
accepted as the price of a protocol that does not depend on hardware monopolies or
unbroken classical cryptography.

Tail emission guarantees that the security budget never reaches zero---not in 2140,
not in 2400, not ever. Every block mined produces at least 0.15\,LAT, ensuring miners
are always compensated and the chain is always defended. Combined with 9 halvings
across $\sim$21 years and a final supply converging toward $\sim$42 million LAT,
the emission curve balances scarcity with perpetual security.

\textbf{What comes next is up to you.} Lattice is live. The software runs on any machine
with a CPU and 2\,GB of RAM. A single Docker command launches a full node. Mining
requires no specialized hardware, no pool registration, no permission. The network is
young, the difficulty is low, and the early blocks are waiting.

Every cryptocurrency faces the same bootstrapping question: why should anyone participate
before the network is valuable? The answer for Lattice is the same answer Satoshi gave
in 2009---because the technology is sound, the timing is right, and the alternative is
to wait until it is too late.

The difference is what Lattice protects against. Bitcoin was built for a world of
classical computers. Lattice is built for what comes after.

Run a node. Mine a block. Read the code. The protocol is open, the mathematics are
published, and the genesis block is already written:

\begin{center}
\emph{``There's plenty of room at the bottom.'' --- R.\ Feynman, 1959}
\end{center}

\noindent The acceleration ($dv/dt$) of proof-of-work drives each block forward,
descending to the subatomic security of lattice mathematics---where room is infinite.

\begin{center}
\Large
$\mathcal{L}$

\normalsize
\textbf{Lattice: A Post-Quantum Settlement Layer.}

\smallskip
\small
\texttt{dt}
\end{center}

%% ============================================================================
%% References
%% ============================================================================

\end{document}